\documentclass[12pt]{article}
\usepackage{axodraw}
\usepackage[dvips]{graphicx}

\def\dspace{\baselineskip = 0.30in}

\def\lapproxeq{\lower .7ex\hbox{$\;\stackrel{\textstyle
<}{\sim}\;$}}
\def\gapproxeq{\lower .7ex\hbox{$\;\stackrel{\textstyle
>}{\sim}\;$}}

\begin{document}

\dspace

\begin{titlepage}
\begin{flushright}
BA-03-24\\
\end{flushright}
\vskip 2cm
\begin{center}
{\Large\bf
Inflation and Leptogenesis \\
~~in Five Dimensional $SO(10)$ 
}
\footnote{
To appear in the Proceedings of 
the 2nd Workshop on ``Physics Beyond The Standard Model'', held  
at Ain Shams Univ., Cairo, Egypt, Feb. 17--20, 2003. 
}
\vskip 1cm
{\normalsize\bf
Bumseok Kyae, and
Qaisar Shafi
}
\vskip 0.5cm
{\it Bartol Research Institute, University of Delaware, \\Newark,
DE~~19716,~~USA\\[0.1truecm]}

%

\end{center}
\vskip 1.5cm


\begin{abstract}

We discuss five dimensional (5D) supersymmetric $SO(10)$ compactified
on the orbifold $S^1/(Z_2\times Z_2')$ such that the $SO(10)$ gauge symmetry
is broken on the fixed point(s) to its maximal subgroup(s).   
The MSSM gauge symmetry is recovered by the usual Higgs mechanism, and 
inflation is associated with the Higgs mechanism, 
implemented through F-term scalar potentials on the two fixed points.
%
%
The spontaneous breaking scale is fixed from $\delta T/T$ measurements
to be around $10^{16}$ GeV, and the scalar spectral index $n=0.98-0.99$.
The inflaton field decays into right-handed neutrinos whose subsequent
out of equilibrium decay yield the observed baryon asymmetry
via leptogenesis.

\end{abstract}
\end{titlepage}

\newpage


\section{Introduction}

Supersymmetric grand unified theories (GUTs) provide an especially attractive
framework for physics beyond the standard model (and MSSM), and
it is therefore natural to ask if there exists in this framework a compelling,
perhaps even an intimate connection with inflation.
In Refs.~\cite{hybrid,hybrid2} one possible approach to this question 
was presented.
In its simplest realization, inflation is associated with the breaking
at scale $M$ of a grand unified gauge group $G$ to $H$.  Indeed, inflation is
`driven' by quantum corrections which arise from the breaking of
supersymmetry (SUSY) by the vacuum energy density in the early universe.
The density fluctuations, it turns out, are proportional to
$(M/M_{\rm Planck})^2$, where $M_{\rm Planck}\simeq 1.2\times 10^{19}$ GeV
denotes the Planck mass.  From the variety of $\delta T/T$ 
measurements~\cite{cobe}, 
especially by the Wilkinson Microwave Anisotropy Probe 
(WMAP)~\cite{wmap,wmap2}, the symmetry breaking scale $M$ is 
of order $10^{16}$ GeV, which is tantalizingly close to 
the grand unification scale 
inferred from the evolution of the minimal supersymmetric standard model 
(MSSM) gauge couplings.  
It is therefore natural to try to realize this inflationary scenario
within a grand unified framework~\cite{hybrid2}.   
Because of the logarithmic radiative corrections that drive inflation,
the spectrum of scalar density fluctuations turns out to be essentially flat.
For the simplest models, the scalar spectral index was found to be
$n_s=0.98(\pm 0.01)$~\cite{hybrid}, in excellent agreement
with a variety of observations~\cite{obs} including the recent WMAP data.
The variation $dn_s/d{\rm ln}k$ of the spectral index is found to be small
($\sim 10^{-3}$).

The $SO(10)$ model~\cite{SO10} is particularly attractive 
in view of the growing confidence
in the existence of neutrino oscillations~\cite{nuoscil},
which require that at least two of the three known neutrinos have a non-zero
mass. Because of the presence of right handed  neutrinos (MSSM singlets),
non-zero masses for the known neutrinos is an automatic consequence
of the see-saw mechanism~\cite{seesaw}.
Furthermore, the right handed neutrinos play an essential role in generating
the observed baryon asymmetry via leptogenesis~\cite{lepto},
which becomes especially compelling within an inflationary
framework~\cite{ls}.
Indeed, an inflationary scenario would be incomplete
without explaining the origin of the observed baryon asymmetry,
and the kind of models we are interested in here
automatically achieve this via leptogenesis.

A realistic supersymmetric inflationary model along
the lines was presented in \cite{khalil},
based on the $SO(10)$ subgroup
$SU(4)_c\times SU(2)_L\times SU(2)_R$~\cite{ps}.
The scalar spectral index $n$ has a value very close to unity
(typically $n\approx 0.98-0.99$),
while the symmetry breaking scale of 
$SU(4)_c\times SU(2)_L\times SU(2)_R$ lies,
as previously indicated, around $10^{16}$ GeV.
The vacuum energy density during inflation is of order $10^{14}$ GeV,
so that the gravitational contribution to the quadrupole anisotropy
is essentially negligible.  It is important to note here that
the inflaton field in this scenario eventually decays into right handed
neutrinos, whose out of equilibrium decays lead to leptogenesis.
An extension to the full $SO(10)$ model is complicated by the notorious
doublet-triplet splitting problem,
which prevents a straightforward implementation of the inflationary scenario.
Of course, the subgroup $SU(4)_c\times SU(2)_L\times SU(2)_R$ neatly evades 
this problem and even allows for
a rather straightforward resolution of the `$\mu$ problem.'

Orbifold symmetry breaking in higher dimensional grand unified theories (GUTs)
have recently attracted a great deal of attention~\cite{Orbifold}.  
There are good reasons for discussing such a breaking mechanism.  
Consider, for instance, the case of $SU(5)$~\cite{SU5} 
in four dimensions (4D).
The presence of dimension five baryon number violating operators
mediated through Higgsino exchange implies
in the `minimal' scheme a proton life time
$\tau_{p\rightarrow K^+\bar{\nu}}\sim 10^{30\pm 2}$ yrs.
This may be in conflict with the recent lower bounds
($\tau_p> 1.9\times 10^{33}$ yrs) for $p\rightarrow K^+\bar{\nu}$
determined by the Superkamiokande experiment~\cite{kamio}.
There are other serious issues such as the notorious doublet-triplet (DT)
splitting problem,
which have led people to investigate five (and higher) dimensional theories
compactified on suitable orbifolds that provide a relatively painless way
of implementing the DT splitting.
Furthermore, dimension five proton decay can be easily eliminated which is
an attractive feature of the five dimensional framework.

In this paper we investigate the construction and implications of
five dimensional (5D) $SO(10)$
compactified on the orbifold $S^1/(Z_2\times Z_2')$ such that
on each of the two fixed points (branes B1 and B2),
the gauge symmetries on the branes correspond
to $SO(10)$ or one of the maximal subgroups of $SO(10)$~\cite{5dso10}.
Thus, after compactification,
the residual 4D gauge symmetry group is
one of the maximal subgroups of $SO(10)$ or
$SU(3)_c\times SU(2)_L\times U(1)_Y\times U(1)_X$.
MSSM gauge symmetry is achieved
via the standard Higgs mechanism.
The MSSM gauge group is realized by spontaneously breaking  
with brane or bulk Higgs. 
Inflation is associated with the Higgs mechanism, followed by its decay
into right-handed neutrinos, which subsequently generate
a primordial lepton asymmetry.
The gravitino constraint on the reheat temperature~\cite{gravitino}
imposes important constraints on the masses of the right-handed neutrinos
which can be folded together with the information now available
from the oscillation experiments~\cite{nuoscil}.

As emphasized in~\cite{KS,ks2}, implementation in five dimension of the
inflationary scenario considered in \cite{hybrid} requires some care.
Note that the 5D setup is the appropriate one because of
the proximity of the scale of inflation and the comapactification scale
(both are of order $M_{GUT}$).
The inflaton potential must be localized on the orbifold fixed points
(branes), since a superpotential in the bulk is not allowed.
For a vanishing bulk cosmological constant, a three space inflationary
solution triggered by non-zero brane tensions (or vacuum energies)
exists~\cite{KS,ks2}.
However, 5D Einstein equations often require that the signs of the brane
tensions on the two ranes are opposite, which  is undesirable.
As shown in~\cite{KS,ks2}, this problem can be circumvented
by introducing a brane-localized
Einstein-Hilbert term in the action.
The two brane tensions are both positive during inflation,
and they vanish when it ends.

The plan of the paper is as follows.
In section 2, we describe the compactification scenario and
the various symmetry breaking patterns from 5D $SO(10)$.
%
%
Section 3 is devoted to the general discussion of inflation and cosmology 
in 5D brane world. 
In section 4, we review the results of 4D F-term inflationary model, 
and discuss leptogenesis.  
In section 5 and 6, we try to embed the 4D inflationary scenario 
in 5D brane setup, and construct realistic models.   
We conclude in section 7.

\section{Orbifold Symmetry Breakings in 5D $SO(10)$}

The $SO(2n)$ generators are represented as
$\left(\begin{array}{cc}
A+C&B+S\\
B-S&A-C
\end{array}\right)$,
where $A$,$B$, $C$ are $n\times n$ anti-symmetric matrices
and $S$ is an $n\times n$ symmetric matrix \cite{zee}.
By an unitary transformation,
the generators are given by
\begin{eqnarray} \label{general}
\left(\begin{array}{cc}
A-iS&C+iB\\
C-iB&A+iS
\end{array}\right)~,
\end{eqnarray}
where $A$ and $S$ denote $U(n)$ generators, and
$C\pm iB$ transform under $SU(n)$ as $n(n-1)/2$ and $\overline{n(n-1)/2}$,
respectively.
Under $SU(5)\times U(1)_X$, the $SO(10)$ generators are decomposed as
\begin{eqnarray} \label{so10}
T_{SO(10)}=\left[\begin{array}{c|c}
{\bf 24}_0+{\bf 1}_0& {\bf 10}_{-4}\\
\hline
{\bf \overline{10}}_{4}&{\bf \overline{24}}_0-{\bf 1}_0
\end{array}\right]_{10\times 10} ~,
\end{eqnarray}
where the subscripts labeling the $SU(5)$ representations indicate
$U(1)_X$ charges, and the subscript ``$10\times 10$'' denotes the matrix
dimension.  Also, ${\bf 24}$ ($={\bf \overline{24}}$) corresponds
to $SU(5)$ generators, while
${\rm diag}~({\bf 1}_{5\times 5},-{\bf 1}_{5\times 5})$
is the $U(1)_X$ generator.
The $5\times 5$ matrices ${\bf 24}_0$ and ${\bf 10}_{-4}$ are further
decomposed under $SU(3)_c\times SU(2)_L\times U(1)_Y$ as
\begin{eqnarray} \label{24}
&&{\bf 24}_{0}=\left(\begin{array}{cc}
{\bf (8,1)}_{0}+{\bf (1,1)}_0 & {\bf (3,\overline{2})}_{-5/6}\\
{\bf (\overline{3},2)}_{5/6} & {\bf (1,3)}_{0}-{\bf (1,1)}_0
\end{array}\right)_{0}~,  \\
&&{\bf 10}_{-4}=\left(\begin{array}{cc}
{\bf (\overline{3},1)}_{-2/3} & {\bf (3,2)}_{1/6}\\
{\bf (3,2)}_{1/6} & {\bf (1,1)}_{1}
\end{array}\right)_{-4} ~.
\end{eqnarray}
%
%
Thus, each representation carries two independent $U(1)$ charges.
Note that the two ${\bf (3,2)}_{1/6}$s in ${\bf 10}_{-4}$ are identified.
%
%

We intend to break $SO(10)$ to its maximal subgroups
by $Z_2$ orbifoldings.
Let us consider the action on $SO(10)$ of the following $Z_2$ group elements,
\begin{eqnarray} \label{p1}
P_1&=&{\rm diag.}\bigg(+I_{3\times 3},+I_{2\times 2},
+I_{3\times 3},+I_{2\times 2}\bigg)~\longrightarrow SO(10)~, \\
P_2&=&{\rm diag.}\bigg(+I_{3\times 3},+I_{2\times 2},
-I_{3\times 3},-I_{2\times 2}\bigg)~\longrightarrow
SU(5)\times U(1)_X, \\
P_3&=&{\rm diag.}\bigg(-I_{3\times 3},+I_{2\times 2},
+I_{3\times 3},-I_{2\times 2}\bigg)~\longrightarrow
SU(5)'\times U(1)_X'~, \\  \label{p4}
P_4&=&{\rm diag.}\bigg(+I_{3\times 3},-I_{2\times 2},
+I_{3\times 3},-I_{2\times 2}\bigg)~\longrightarrow
SU(4)_c\times SU(2)_L\times SU(2)_R ~,   ~~~
\end{eqnarray}
where $I$'s denote identity matrices.
Here the $P$'s all satisfy $P^2=I_{5\times 5}$.
Eqs.~(\ref{p1})--(\ref{p4}) show all possible ways to define
the 10 dimensional $Z_2$ group elements and the maximal subgroups
of $SO(10)$ obtained by their operations, as will be explained below.

Under the operations $P_1T_{SO(10)}P_1^{-1}$, $P_2T_{SO(10)}P_2^{-1}$,
$\cdots$, the matrix elements of $T_{SO(10)}$ transform as
\begin{eqnarray} \label{so10/z2z2}
\left[\begin{array}{cc|cc}
{\bf (8,1)}_{0}^{++++} &
~{\bf (3,\overline{2})}_{-5/6}^{++--} &
~{\bf (\overline{3},1)}_{-2/3}^{+--+} & {\bf (3,2)}_{1/6}^{+-+-} \\
{\bf (\overline{3},2)}_{5/6}^{++--} &
~{\bf (1,3)}_{0}^{++++} &
~{\bf (3,2)}_{1/6}^{+-+-} & {\bf (1,1)}_{1}^{+--+} \\
\hline
{\bf (3,1)}_{2/3}^{+--+} & ~{\bf (\overline{3},\overline{2})}_{-1/6}^{+-+-} &
~{\bf (8,1)}_{0}^{++++}
& {\bf (\overline{3},2)}_{5/6}^{++--} \\
{\bf (\overline{3},\overline{2})}_{-1/6}^{+-+-} & ~{\bf (1,1)}_{-1}^{+--+} &
~{\bf (3,\overline{2})}_{-5/6}^{++--} &
{\bf (1,3)}_{0}^{++++}
\end{array}\right]_{10\times 10} ~,
\end{eqnarray}
where the superscripts of the matrix elements indicate the eigenvalues
of $P_1$, $P_2$, $P_3$, and $P_4$ respectively.
Here, to avoid too much clutter, we have omitted the two $U(1)$ generators
(${\bf (1,1)_0^{++++}}$).
As shown in Eqs.~(\ref{so10}) and (\ref{24}), they appear
in the diagonal part of the matrix (\ref{so10/z2z2}).

For future convenience, let us define the $SO(10)$ generator pieces
appearing in Eq.~(\ref{so10/z2z2}) more succinctly,
\begin{eqnarray} \label{su5}
\left[\begin{array}{cc|cc}
{\bf G}^{++++} & ~{\bf Q'}^{++--} &
~{\bf U^{c}}^{+--+} & {\bf Q}^{+-+-} \\
~{\bf \overline{Q'}}^{++--} & ~{\bf W}^{++++} &
{\bf Q}^{+-+-} & ~{\bf E^{c}}^{+--+} \\
\hline
~{\bf \overline{U}^{c}}^{+--+} & {\bf \overline{Q}}^{+-+-} &
{\bf G}^{++++} & ~{\bf \overline{Q'}}^{++--} \\
{\bf \overline{Q}}^{+-+-} & ~{\bf \overline{E}^{c}}^{+--+} &
~{\bf Q'}^{++--} & ~{\bf W}^{++++}
\end{array}\right] ~,
\end{eqnarray}
whose entries are in one to one correspondence
to those of Eq.~(\ref{so10/z2z2}).
Note that ${\bf Q}$ denotes ${\bf (3,2)}_{1/6}$, while ${\bf Q'}$ denotes
${\bf (3,\overline{2})}_{-5/6}$.
Similarly, the two $U(1)$ generators ${\bf (1,1)_0^{++++}}$,
which were omitted in Eq.~(\ref{so10/z2z2}), are defined as
\begin{eqnarray}
{\bf Y}^{++++}~~~ {\rm and}~~~ {\bf X}^{++++}~,
\end{eqnarray}
where ${\bf Y}$ corresponds to the hypercharge generator of SM.
We identify the eigenvalues of the above generators with those of
the associated gauge fields (and gauginos).

Suppose we have an $S^1/(Z_2\times Z_2')$ orbifold compactification
in 5D space-time.
The two $Z_2$ elements among Eqs.~(\ref{p1})--(\ref{p4})
can be employed so as to embed the internal $Z_2\times Z_2'$ into
the two presumed reflection symmetries for the extra space,
$y\leftrightarrow -y$ and $y'\leftrightarrow -y'$ ($y'=y+y_c/2$).
Two eigenvalues of $P_{i}$ could be interpreted as
the parities (or boundary conditions) of the relevant fields
under such reflections~\cite{hebecker}.
Thus, the wave function of a field with parity $(+-)$, for instance,
must vanish on the brane at $y=y_c/2$ (B2),
while it survives at $y=0$ brane (B1).  Only those fields assigned $(++)$
parities contain massless modes in their Kaluza-Klein (KK) spectrum.
Thus, even though the bulk Lagrangian respects $SO(10)$,
the effective low energy theory possesses a smaller gauge symmetry
associated with the $(++)$ generators.

If $P_1$ (identity) and one more $P_i$ ($i=2,3,4$) are taken
as $Z_2\times Z_2'$ elements,
the $SO(10)$ gauge symmetry breaks to
$SU(5)\times U(1)_X$, $SU(5)'\times U(1)_X'$~\cite{5dflipped}, and
$SU(4)_c\times SU(2)_L\times SU(2)_R$~\cite{5dps,KS,hdkim,5dwmap}, 
respectively.
On the other hand, with two different $P_i$'s from among $\{P_2, P_3, P_4\}$,
$SO(10)$ can be broken to
$SU(3)_c\times SU(2)_L\times U(1)_Y\times U(1)_X$~\cite{ks2,hebecker2},
as illustrated in Figure 1.
\begin{figure}
\begin{center}
\begin{picture}(225,225)(0,0)


\CArc(112.5,131.25)(52.5,108,72)
\CArc(90,92.25)(52.5,182,167)
\CArc(135,92.25)(52.5,15,359)
\Text(112.5,161)[]{${\bf Q'}$, ${\bf \overline{Q'}}$}
\Text(62,78)[]{${\bf Q}$, ${\bf \overline{Q}}$}
\Text(163.5,82.5)[]{${\bf U^c}$, ${\bf \overline{U}^c}$}
\Text(163.5,71.25)[]{${\bf E^c}$, ${\bf \overline{E}^c}$}
\Text(112.5,114)[]{${\bf SM}$}
\Text(112.5,100)[]{${\bf \times U(1)_X}$}
\Text(112.5,183.75)[]{${\bf 5-1}$}
\Text(38.25,99)[]{${\bf 5'-1'}$}
\Text(187.5,99)[]{${\bf 4-2-2}$}
\Text(112.5,210)[]{${\bf SO(10)}$}
\Line(89,210)(7.5,210)
\Line(7.5,210)(7.5,15)
\Line(7.5,15)(220,15)
\Line(220,15)(220,210)
\Line(220,210)(135,210)


\end{picture}
\caption{A diagram showing the generators of $SO(10)$ and its subgroups.
$5-1$, $5'-1'$, $4-2-2$, and SM denote $SU(5)\times U(1)_X$,
$SU(5)'\times U(1)_X'$, $SU(4)_c\times SU(2)_L\times SU(2)_R$,
and the MSSM gauge group, respectively.}
\end{center}
\end{figure}
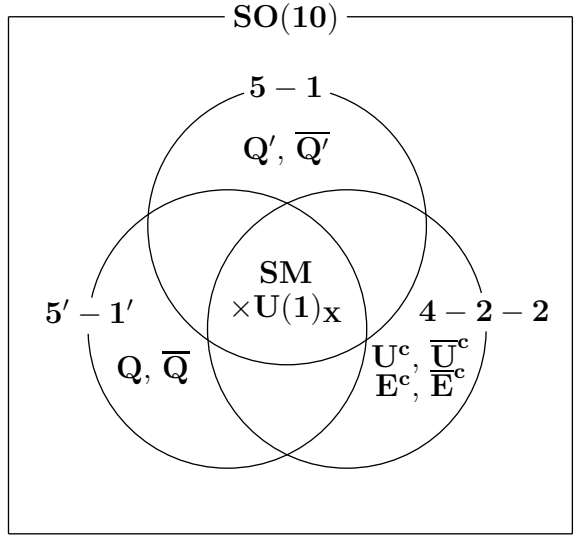

As is well known, compactification on $S^1/(Z_2\times Z_2')$ also
can break the 4D $N=2$ SUSY to $N=1$.
An $N=2$ SUSY vector multiplet is split into
an $N=1$ vector multiplet and a chiral multiplet in adjoint representation.
We assign the same parities as the generators
to the associated vector multiplets as claimed above,
but opposite parities to chiral multiplets.
Then, $N=2$ SUSY is broken to $N=1$.

\subsection{$SO(10)$~--~$SU(4)_c\times SU(2)_L\times SU(2)_R$}

An effective 4D theory with the gauge group 
$SU(4)_c\times SU(2)_L\times SU(2)_R$ is readily obtained
from a 5D $SO(10)$ gauge theory with $P_1$ and $P_4$,   
if the fifth dimension is compactified
on the orbifold $S^1/(Z_2\times Z_2')$ \cite{5dps}, where
$Z_2$ reflects $y\rightarrow -y$, and $Z_2'$ reflects $y'\rightarrow -y'$
with $y'=y+y_c/2$.
There are two independent orbifold fixed points (branes)
at $y=0$ and $y=y_c/2$, with $N=1$ SUSYs and gauge symmetries $SO(10)$
and $SU(4)_c\times SU(2)_L\times SU(2)_R$ respectively \cite{5dps}.
The $SO(10)$ gauge multiplet $(A_M,\lambda^1,\lambda^2,\Phi)$
decomposes under $SU(4)_c\times SU(2)_L\times SU(2)_R$ as
\begin{eqnarray}
V_{\bf 45}&\longrightarrow& V_{({\bf 15},{\bf 1},{\bf 1})}+
V_{({\bf 1},{\bf 3},{\bf 1})}+V_{({\bf 1},{\bf 1},{\bf 3})}
+V_{({\bf 6},{\bf 2},{\bf 2})} \\
&&+\Sigma_{({\bf 15},{\bf 1},{\bf 1})}
+\Sigma_{({\bf 1},{\bf 3},{\bf 1})}+\Sigma_{({\bf 1},{\bf 1},{\bf 3})}
+\Sigma_{({\bf 6},{\bf 2},{\bf 2})} ~,  \nonumber
\end{eqnarray}
where $V$ and $\Sigma$ denote the vector multiplet $(A_\mu,\lambda^1)$ and
the chiral multiplet $((\Phi+iA_5)/\sqrt{2},\lambda^2)$ respectively.  
Here, $({\bf 15},{\bf 1},{\bf 1})+({\bf 1},{\bf 3},{\bf 1})+
({\bf 1},{\bf 1},{\bf 3})$ corresponds to $SU(4)_c\times SU(2)_L\times SU(2)_R$
generators composed of  
\begin{eqnarray}
({\bf 15},{\bf 1},{\bf 1})+({\bf 1},{\bf 3},{\bf 1})+({\bf 1},{\bf 1},{\bf 3})
={\bf G}+{\bf U^c}+{\bf \overline{U}^c}+{\bf X}+{\bf W}+{\bf E^c}
+{\bf \overline{E}^c}+{\bf Y} ~,  
\end{eqnarray}
while $({\bf 6},{\bf 2},{\bf 2})$ consists of 
\begin{eqnarray}
({\bf 6},{\bf 2},{\bf 2})={\bf Q}+{\bf \overline{Q}}
+{\bf Q'}+{\bf \overline{Q'}} ~.
\end{eqnarray} 
$(Z_2,Z_2')$ parity assignments and KK masses of $V$'s and $\Sigma$'s 
are shown in Table I.
\vskip 1cm
\begin{center}
\begin{tabular}{|c||c|c|c|c|} \hline
Vector &$V_{({\bf 15},{\bf 1},{\bf 1})}$ & $V_{({\bf 1},{\bf 3},{\bf 1})}$ &
$V_{({\bf 1},{\bf 1},{\bf 3})}$ &$V_{({\bf 6},{\bf 2},{\bf 2})}$ \\
\hline \hline
$(Z_2,Z_2')$ &(+,+)$$ &$(+,+)$ &$(+,+)$ &$(+,-)$ \\
Masses & $2n\pi/y_c$ & $2n\pi/y_c$ & $2n\pi/y_c$ & $(2n+1)\pi/y_c$ \\
\hline \hline
Chiral & $\Sigma_{({\bf 15},{\bf 1},{\bf 1})}$ &
$\Sigma_{({\bf 1},{\bf 3},{\bf 1})}$ &
$\Sigma_{({\bf 1},{\bf 1},{\bf 3})}$ &$\Sigma_{({\bf 6},{\bf 2},{\bf 2})}$
\\ \hline \hline
$(Z_2,Z_2')$ & $(-,-)$ &$(-,-)$ &$(-,-)$ &$(-,+)$  \\
Masses & $(2n+2)\pi/y_c$ & $(2n+2)\pi/y_c$ & $(2n+2)\pi/y_c$ & $(2n+1)\pi/y_c$
\\
\hline
\end{tabular}
\end{center}
{\bf Table I.~}($Z_2,Z_2'$) parity assignments and
Kaluza-Klein masses $(n=0,1,2,\cdots)$ ~~~~~~~~~~~~
\hspace{2.0cm} for the vector multiplet in $SO(10)$.
\vskip 0.2cm

The parities of the chiral multiplets $\Sigma$'s are opposite to those of
the vector multiplets $V$'s in Table I and
hence, $N=2$ SUSY explicitly breaks to $N=1$
below the compactification scale $\pi/y_c$.
As shown in Table I, only the vector multiplets,
$V_{({\bf 15},{\bf 1},{\bf 1})}$, $V_{({\bf 1},{\bf 3},{\bf 1})}$, and
$V_{({\bf 1},{\bf 1},{\bf 3})}$ contain massless modes, which
means that the low energy effective 4D theory reduces to $N=1$
supersymmetric $SU(4)_c\times SU(2)_L\times SU(2)_R$.
The parity assignments in Table I also show that the wave function
of the vector multiplet $V_{({\bf 6},{\bf 2},{\bf 2})}$ vanishes
at the brane located at $y=y_c/2$ (B2) because it is assigned an odd parity
under $Z_2'$, while the wave functions of all the vector multiplets should
be the same at the $y=0$ brane (B1).
Therefore, while the gauge symmetry at B1 is $SO(10)$,
only $SU(4)_c\times SU(2)_L\times SU(2)_R$ is preserved at B2 \cite{hebecker}.

\subsection{$SU(5)\times U(1)_X$~--~$SU(5)'\times U(1)_X'$}

Let us consider the case in which $P_2$ and $P_3$ operations
are chosen as $Z_2$ and $Z_2'$ elements~\cite{ks2},
corresponding to the second and the third parities
in Eq.~(\ref{so10/z2z2}) and (\ref{su5}).
With $P_2$, positive parities are assigned to the block-diagonal
elements ($SU(5)\times U(1)_X$ generators and
their associated gauge multiplets),
while with $P_3$, positive parities are assigned to the generators of
$SU(3)_c\times SU(2)_L\times U(1)_Y\times U(1)_X$
and ${\bf Q}^{-+}$, ${\bf \overline{Q}}^{-+}$, and
to their associated gauge multiplets.
Hence, after compactification, the gauge symmetry reduces
to $SU(3)_c\times SU(2)_L\times U(1)_Y\times U(1)_X$.
Together with ${\bf X}^{++}_V$, the $SU(5)$ gauge multiplets
in Eq.~(\ref{su5}) survive at B1,
\begin{eqnarray} \label{24}
{\bf 24}_V=\bigg({\bf G}_V^{++}+{\bf W}_V^{++}+{\bf Y}_V^{++}\bigg)
+\bigg({\bf Q'}_V^{+-}+{\bf \overline{Q'}}_V^{+-}\bigg)~~~
{\rm at~B1}~,
\end{eqnarray}
where the subscripts $V$ denote the vector multiplets.
Thus $SU(5)\times U(1)_X$ should be preserved at B1~\cite{hebecker}.

At B2 ${\bf Q'}_V^{+-}$ and ${\bf \overline{Q'}}_V^{+-}$ in Eq.~(\ref{24})
are replaced by ${\bf Q}_V^{-+}$ and ${\bf \overline{Q}}_V^{-+}$,
which are in the ${\bf 10}_{-4}$ and ${\bf \overline{10}}_{4}$
representations of $SU(5)\times U(1)_X$,
\begin{eqnarray}
{\bf 24}_V'=\bigg({\bf G}_V^{++}+{\bf W}_V^{++}+{\bf Y}_V^{++}\bigg)
+\bigg({\bf Q}_V^{-+}+{\bf \overline{Q}}_V^{-+}\bigg)~~~
{\rm at~B2}~.
\end{eqnarray}
Note that the assigned hypercharges coincide
with those given in `flipped' $SU(5)'\times U(1)_X'$~\cite{flipped}.
The $U(1)_X'$ generator at B2, ${\bf X'}^{++}$ is defined as
\begin{eqnarray}
{\rm diag}({\bf 1}_{3\times 3}, -{\bf 1}_{2\times 2},
-{\bf 1}_{3\times 3}, {\bf 1}_{2\times 2}) ~.
\end{eqnarray}
Thus, the $U(1)_X'$ charges of the surviving elements at B2
turn out to be zero, while the other components are
assigned $-4$ or $4$.
The $U(1)_X'$ generator and the matrix elements with $(++)$, $(-+)$ parities
in Eq.~(\ref{so10/z2z2})
can be block-diagonalized to the form given in Eq.~(\ref{so10})
\begin{eqnarray} \label{flipped}
\left[\begin{array}{cc|cc}
{\bf G}^{++} & ~{\bf Q}^{-+} &
~{\bf U^{c}}^{--} & {\bf Q'}^{+-} \\
{\bf \overline{Q}}^{-+} & ~{\bf W}^{++} &
{\bf Q'}^{+-} & ~{\bf \overline{E}^{c}}^{--} \\
\hline
~{\bf \overline{U}^{c}}^{--} & ~{\bf \overline{Q'}}^{+-} &
{\bf G}^{++} & ~{\bf \overline{Q}}^{-+} \\
{\bf \overline{Q'}}^{+-} & ~{\bf E^{c}}^{--} &
{\bf Q}^{-+} & ~{\bf W}^{++}
\end{array}\right] ~,
\end{eqnarray}
through unitary transformation
of the $SO(10)$ generator in Eq.~(\ref{so10/z2z2}) with
\begin{eqnarray}
U_3=\left(\begin{array}{c|ccc}
I_{3\times 3} & 0 & 0 & 0 \\ \hline
0 & 0 & 0 & I_{2\times 2} \\
0 & 0 & I_{3\times 3} & 0 \\
0 & I_{2\times 2} & 0 & 0
\end{array}\right)_{10\times 10} ~.
\end{eqnarray}
In Eq.~(\ref{flipped}), the two superscripts denote the eigenvalues
of $P_2$ and $P_3$.
From Eq.~(\ref{flipped}), we conclude that
the gauge symmetry at B2
is associated with a second (flipped) $SU(5)'\times U(1)_X'$
embedded in $SO(10)$~\cite{flipped}.
%

To break 4D $N=2$ SUSY, opposite parities should be assigned to
the chiral multiplet $(\Phi+iA_5,\lambda_2)$, where $\Phi$, $A_5$, $\lambda_2$
belong to $N=2$ vector multiplets.
The non-vanishing chiral multiplets at B1 are
\begin{eqnarray} \label{chiral1}
{\bf 10}_\Sigma &=& {\bf U^c}_\Sigma^{++}+{\bf E^c}_\Sigma^{++}
+{\bf Q}_\Sigma^{+-} ~,~  \\ \label{chiral1'}
{\bf \overline{10}}_\Sigma &=& {\bf \overline{U}^c}_\Sigma^{++}
+{\bf \overline{E}^c}_\Sigma^{++}
+{\bf \overline{Q}}_\Sigma^{+-} ~~~~ ~,
\end{eqnarray}
while on B2, ${\bf Q}_\Sigma^{+-}$ and ${\bf \overline{Q}}_\Sigma^{+-}$
are replaced by
${\bf Q'}_\Sigma^{+-}$ and ${\bf \overline{Q'}}_\Sigma^{+-}$
(which are in ${\bf 24}_\Sigma$ and ${\bf 24'}_\Sigma$ at B1).
Together with the vector-like pairs with $(++)$ parities,
they compose ${\bf 10}_{-4}'$ and ${\bf \overline{10}_4}'$-plets
of $SU(5)'\times U(1)_X'$ at B2,
\begin{eqnarray}
{\bf 10'}_\Sigma &=& {\bf U^c}_\Sigma^{++}+{\bf \overline{E}^c}_\Sigma^{++}
+{\bf Q'}_\Sigma^{-+} ~~,~  \\
{\bf \overline{10'}}_\Sigma &=& {\bf \overline{U}^c}_\Sigma^{++}
+{\bf E^c}_\Sigma^{++}
+{\bf \overline{Q'}}_\Sigma^{-+} ~. \label{chiral2}
\end{eqnarray}
We note in Eqs.~(\ref{chiral1})--(\ref{chiral2}) the appearance of
two vector-like pairs
${\bf U^c}_\Sigma^{++}$, ${\bf \overline{U}^c}_\Sigma^{++}$ and
${\bf E^c}_\Sigma^{++}$, ${\bf \overline{E}^c}_\Sigma^{++}$,
which contain massless modes.
We summarize the above results in Table II.
\vskip 0.4cm
\begin{center}
\begin{tabular}{|c|c||cccccc|} \hline
Vector (B1) ~&~ ${\bf 24}_V$, ${\bf 1}_V$ ~&~
${\bf G}_V^{++}$~, & ${\bf W}_V^{++}$~, & ${\bf Y}_V^{++}$~, &
${\bf X}_V^{++}$~, & ${\bf Q'}_V^{+-}$~, & ${\bf Q'}_V^{+-}$~
\\
Chiral (B1) ~&~ ${\bf 10}_\Sigma$, ${\bf \overline{10}}_\Sigma$ ~&~
${\bf U^c}_\Sigma^{++}$, & ${\bf E^c}_\Sigma^{++}$, &
${\bf Q}_\Sigma^{+-}$~, & ${\bf \overline{U}^c}_\Sigma^{++}$, &
${\bf \overline{E}^c}_\Sigma^{++}$, & ${\bf \overline{Q}}_\Sigma^{+-}$~
\\ \hline \hline
Vector (B2) & ${\bf 24'}_V$, ${\bf 1'}_V$ &~
${\bf G}_V^{++}$~, & ${\bf W}_V^{++}$~, & ${\bf Y}_V^{++}$~, &
${\bf X'}_V^{++}$, & ${\bf Q}^{-+}_V$~, & ${\bf \overline{Q}}_V^{-+}$~
\\
Chiral (B2) & ${\bf 10'}_\Sigma$, ${\bf \overline{10}'}_\Sigma$ &~
${\bf U^c}_\Sigma^{++}$, & ${\bf \overline{E}^c}_\Sigma^{++}$, &
${\bf Q'}_\Sigma^{-+}$~, & ${\bf \overline{U}^c}_\Sigma^{++}$, &
${\bf E^c}_\Sigma^{++}$, & ${\bf \overline{Q'}}_\Sigma^{-+}$~
\\ \hline
\end{tabular}
\vskip 0.4cm
{\bf Table II.~} Surviving superfields on each brane
in the $SO(10)$ gauge multiplet.
\end{center}
To preserve the successful MSSM gauge coupling unification,
we need to remove them from the low energy spectrum.
To realize the MSSM gauge symmetry at lower energies,
we employ the Higgs mechanism via bulk Higgs fields.
This is because with brane Higgs fields, it is hard to provide heavy masses
for the vector-like pairs,
${\bf U^c}_\Sigma^{++}$, ${\bf \overline{U}^c}_\Sigma^{++}$, and
${\bf E^c}_\Sigma^{++}$, ${\bf \overline{E}^c}_\Sigma^{++}$.
Let us introduce two pairs of Higgs hypermultiplets
${\bf 16}$, ${\bf \overline{16}}$ as shown in Table III.
\vskip 0.4cm
\begin{center}
\begin{tabular}{|c||c|c|} \hline
Hypermultiplets & $Z_2\times Z_2'$~ parities & $U(1)_R$
\\ \hline \hline
${\bf 16}_H~$ & ~${\bf u^c}^{--},~{\bf e^c}^{--},~{\bf q}^{-+}~~;~~
{\bf d^c}^{++},~{\bf l}^{+-}~~;~~{\bf \nu^c}^{++}$ & $0$
\\
${\bf 16^c}_H$ & ${\bf u}^{++}~,~{\bf e}^{++}~,~{\bf q^c}^{+-}~;~~
{\bf d}^{--}~,~{\bf l^c}^{-+}~;~~{\bf \nu}^{--}$ & $0$
\\ \hline
${\bf \overline{16}}_H$ & ~${\bf \overline{u}^c}^{--},
~{\bf \overline{e}^c}^{--},~{\bf \overline{q}}^{-+}~~;~~
{\bf \overline{d}^c}^{++},~{\bf \overline{l}}^{+-}~~;~~
{\bf \overline{\nu}^c}^{++}$ &$0$
\\
${\bf \overline{16}^c}_H$ & ${\bf \overline{u}}^{++}~,
~{\bf \overline{e}}^{++}~,~{\bf \overline{q}^c}^{+-}~;~~
{\bf \overline{d}}^{--}~,~{\bf \overline{l}^c}^{-+}~;~~
{\bf \overline{\nu}}^{--}$ & $0$
\\ \hline
\end{tabular}
\vskip 0.4cm
{\bf Table III.~} $Z_2\times Z_2'$ parities of the bulk Higgs hypermultiplets.
\end{center}
From ${\bf 16}_H$ and ${\bf \overline{16}}_H$,
the surviving fields at B1 and B2 are
\begin{eqnarray}
{\bf 16}_H&:&({\bf d^c}^{++},~ {\bf l}^{+-};~ {\bf \nu^c}^{++})
~~~{\rm at~~B1} ~, \\
 &&({\bf d^c}^{++},~{\bf q}^{-+},~{\bf \nu^c}^{++}) ~~~{\rm at~~B2} ~, \\
{\bf \overline{16}}_H&:&({\bf \overline{d}^c}^{++},~ {\bf \overline{l}}^{+-};~
{\bf \overline{\nu}^c}^{++}) ~~~{\rm at~~B1}~, \\
&&({\bf \overline{d}^c}^{++},{\bf \overline{q}}^{-+},
~{\bf \overline{\nu}^c}^{++})~~~~{\rm at~~B2} ~,
\end{eqnarray}
They compose (${\bf \overline{5}}_{-3}$; ${\bf 1}_{5}$) and
(${\bf 5}_{3}$; ${\bf \overline{1}}_{-5}$) representations
of $SU(5)\times U(1)_X$
at B1, and ${\bf 10}_{1}'$, ${\bf \overline{10}}_{-1}'$
of $SU(5)'\times U(1)_X'$ at B2.

In order to realize $N=1$ SUSY, the surviving fields from
${\bf 16^c}$, ${\bf \overline{16}^c}$ on the two branes should be
as follows:
\begin{eqnarray}
{\bf 16^c}_H&:&({\bf u}^{++},~ {\bf q^c}^{+-},~ {\bf e}^{++})
~~~{\rm at~~B1} ~, \\
&:& ({\bf u}^{++},~{\bf l^c}^{-+};~{\bf e}^{++}) ~~~{\rm at~~B2} ~, \\
{\bf \overline{16}^c}_H&:&({\bf \overline{u}}^{++},~{\bf \overline{q}^c}^{+-},~
{\bf \overline{e}}^{++})~~~{\rm at~~B1} ~, \\
&:& ({\bf \overline{u}}^{++},~{\bf \overline{l}^c}^{-+};
~{\bf \overline{e}}^{++}) ~~~{\rm at~~B2} ~.
\end{eqnarray}
They compose ${\bf 10^c}_{-1}$,
${\bf \overline{10}^c}_{1}$ ($={\bf 10}_{1}$) at B1,  and
(${\bf \overline{5}^{c'}}_{3}$; ${\bf 1^{c'}}_{-5}$),
(${\bf 5^{c'}}_{-3}$; ${\bf \overline{1}^c}_{5}$) at B2,
respectively.
The results are summarized in Table IV.
\vskip 0.4cm
\begin{center}
\begin{tabular}{|c|c|c|c|c|} \hline
B1 & ${\bf 5}_H$, ${\bf 1}_H$ &
${\bf \overline{5}}_H$, ${\bf \overline{1}}_H$ & ${\bf 10^c}_H$ &
${\bf \overline{10}^c}_H$
\\ \hline
 & ${\bf d^c}^{++}$, ${\bf l}^{+-}$, ${\bf \nu^c}^{++}$ &
${\bf \overline{d}^c}^{++}$, ${\bf \overline{l}}^{+-}$,
${\bf \overline{\nu}^c}^{++}$ &
${\bf u}^{++}$, ${\bf e}^{++}$, ${\bf q^c}^{+-}$ &
${\bf \overline{u}}^{++}$, ${\bf \overline{e}}^{++}$,
${\bf \overline{q}^c}^{+-}$
\\ \hline \hline
B2 & ${\bf 10'}_H$ & ${\bf \overline{10}'}_H$ &
${\bf 5^{'c}}_H$, ${\bf 1^{'c}}_H$ &
${\bf \overline{5}^{'c}}_H$, ${\bf \overline{1}^{'c}}_H$
\\ \hline
 & ${\bf d^c}^{++}$, ${\bf \nu^c}^{++}$, ${\bf q}^{-+}$ &
${\bf \overline{d}^c}^{++}$, ${\bf \overline{\nu}^c}^{++}$,
${\bf \overline{q}}^{-+}$ &
${\bf u}^{++}$, ${\bf l^c}^{-+}$, ${\bf e}^{++}$  &
${\bf \overline{u}}^{++}$, ${\bf \overline{l}^c}^{-+}$,
${\bf \overline{e}}^{++}$
\\ \hline
\end{tabular}
\vskip 0.4cm
{\bf Table IV.~} Surviving Higgs superfields on the branes B1 and B2.
\end{center}

Now let us discuss the $N=2$ (bulk) hypermultiplet $H$ $(=(\phi,\psi))$,
$H^c$ ($=(\phi^c,\psi^c)$) in the vector representations ${\bf 10}$,
${\bf 10^c}$ ($={\bf 10}$) of $SO(10)$, where
$H$ and $H^c$ are $N=1$ chiral multiplets.
Under $SU(5)\times U(1)_X$ and $SU(3)_c\times SU(2)_L\times U(1)_Y$,
${\bf 10}$ and ${\bf 10^c}$ are
\begin{eqnarray}
{\bf 10}=
\left(\begin{array}{c}
{\bf 5}_{-2} \\
\hline
{\bf \overline{5}}_{2}
\end{array}\right)=
\left(\begin{array}{c}
~{\bf (3,1)}_{-1/3}^{+-}  \\
{\bf (1,2)}_{1/2}^{++}  \\
\hline
{\bf (\overline{3},1)}_{1/3}^{-+}  \\
~{\bf (1,\overline{2})}_{-1/2}^{--}
\end{array}\right)~, ~~
{\bf 10^c}=
\left(\begin{array}{c}
{\bf 5^c}_{2} \\
\hline
{\bf \overline{5^c}}_{-2}
\end{array}\right)=
\left(\begin{array}{c}
{\bf (3^c,1)}_{1/3}^{-+}  \\
~{\bf (1,2^c)}_{-1/2}^{--}  \\
\hline
~{\bf (\overline{3}^c,1)}_{-1/3}^{+-}  \\
{\bf (1,\overline{2}^c)}_{1/2}^{++}
\end{array}\right) ~,
\end{eqnarray}
where the subscripts $\pm 2$ are $U(1)_X$ charges and the remaining subscripts
indicate the hypercharges $Y$.
The superscripts on the matrix elements denote the eigenvalues of
the $P$ and $P'$ operations.
As in the $N=2$ vector multiplet,
opposite parities must be assigned for $H$ and $H^c$
to break $N=2$ SUSY to $N=1$.   The massless modes are contained
in the two doublets ${\bf (1,2)}_{1/2}^{++}$ and
${\bf (1,\overline{2^c})}_{1/2}^{++}$ ($={\bf (1,2)}_{1/2}^{++}$).
While the surviving representations at B1,
${\bf (3,1)}_{-1/3}^{+-}$ and ${\bf (1,2)}_{1/2}^{++}$
(also ${\bf (\overline{3}^c,1)}_{-1/3}^{+-}$ and
${\bf (1,\overline{2}^c)}_{1/2}^{++}$) compose two ${\bf 5}_{-2}$
(or ${\bf \overline{5}^c}_{-2}$) of $SU(5)\times U(1)_X$,
at B2  the non-vanishing representations are
two ${\bf \overline{5}}_{2}'$ of $SU(5)'\times U(1)_X'$,
\begin{eqnarray}
{\bf \overline{5}}_{2}' ={\bf (\overline{3},1)}_{1/3}^{-+}
+{\bf (1,2)}_{1/2}^{++} ~~\bigg({\rm or}~~{\bf (3^c,1)}_{1/3}^{-+}
+{\bf (1,\overline{2}^c)}_{1/2}^{++}\bigg) ~~{\rm at~B2}~.
\end{eqnarray}

In this model, the $SU(2)$ $R$-symmetry which generally exists in $N=2$
SUSY theories is explicitly broken to $U(1)_R$.
Since $N=1$ SUSY is present on both branes,
$U(1)_R$ symmetry should be respected.
We note that different $U(1)_R$ charges can be assigned
to $H_{{\bf 10}_{-4}}$ and $H^c_{{\bf 10^c}_{4}}$
as shown in Table V~\cite{nomura}.
\vskip 0.6cm
\vskip 0.6cm
\begin{center}
\begin{tabular}{|c||ccc|c|} \hline
$U(1)_R$&  & $V$,~ $\Sigma$ & & $H$, $H^c$
\\
\hline \hline
$1$ & & & & $\phi^c$
\\
$1/2$ & & $\lambda^1$& & $\psi^c$
\\
$0$ & ~~$A_\mu$& & $\Phi$, $A_5$~~ & $\phi$
\\
$-1/2$ & & $\lambda^2$& & $\psi$
\\
\hline
\end{tabular}
\vskip 0.4cm
{\bf Table V.~}$U(1)_R$ charges of the vector and hypermultiplets.
\end{center}
The results of Table V are consistent with our choice of
the $U(1)_R$ charges $1/2$ ($-1/2$) and $-1/2$ ($1/2$) for the SUSY parameters
$\theta^1$ ($d\theta^1$) and $\theta^2$ ($d\theta^2$), respectively.

Consider the following Higgs superpotentials on the two branes
\begin{eqnarray} \label{sp1}
W_{B1}&=&\kappa_1S\bigg({\bf 16}_H{\bf \overline{16}}_H-M_1^2\bigg)~,
\\ \label{sp2}
W_{B2}&=&\kappa_2S\bigg({\bf 16'}_H{\bf \overline{16}'}_H+{\bf 1}{\bf 1'}
-M_2^2\bigg) ~,
\end{eqnarray}
where $\kappa_{1,2}$ ($M_{1,2}$) are dimensionless (dimensionful) parameters.
Here ${\bf 16}_H{\bf \overline{16}}_H$
stands for the superpotential couplings
by the surviving Higgs at B1 shown in Table IV,
${\bf 10^c}_{H}{\bf \overline{10}^c}_H+{\bf \overline{5}}_{H}{\bf 5}_{H}
+{\bf 1}_H{\bf \overline{1}}_{H}$
with arbitrary coefficients.
${\bf 16'}_H{\bf \overline{16}'}_H$ in Eq.~(\ref{sp2})
is also similarly understood.
$S$ is a bulk singlet superfield with unit $U(1)_R$ charge,
which can couple to the Higgs fields on both branes.
Also, ${\bf 1}$, ${\bf 1'}$ are gauge singlet fields
with suitable $U(1)_R$ charges.
With non-zero vacuum expectation values (VEVs) of the scalar components
of ${\bf \nu^c}^{++}$, ${\bf \overline{\nu}^c}^{c++}$,
$SU(3)_c\times SU(2)_L\times U(1)_Y\times U(1)_X$ is spontaneously broken
to the MSSM gauge group.
Note that suitable VEVs of ${\bf 1}$, ${\bf 1'}$ can ensure
that the VEVs $\langle{\bf \nu^c}^{++}\rangle$ and
$\langle{\bf \overline{\nu}^c}^{c++}\rangle$ are constant
along the extra dimension.

With spontaneous symmetry breaking, the gauge bosons, gauge scalars and
their superpartners in ${\bf 10}_{-4}$, ${\bf \overline{10}}_{4}$
acquire masses.
The gauge bosons in ${\bf U^c}^{--}_V$,
${\bf Q}^{-+}_V$, ${\bf E^c}^{--}_V$, and
${\bf \overline{U}^c}^{--}_V$,
${\bf \overline{Q}}^{-+}_V$, ${\bf \overline{E}^c}^{--}_V$
absorb a linear combination of $A_5$'s from
\begin{eqnarray}
&&{\bf U^c}^{++}_\Sigma~ (n\neq 0)~,~~
{\bf Q}^{+-}_\Sigma~,~~{\bf E^c}^{++}_\Sigma~ (n\neq 0)~,~~~~{\rm and}
\\
&&{\bf \overline{U}^c}^{++}_\Sigma~ (n\neq 0)~,~~
{\bf \overline{Q}}^{+-}_\Sigma~,~~{\bf \overline{E}^c}^{++}_\Sigma~ (n\neq 0)~,
\end{eqnarray}
and from the Higgs fields
\begin{eqnarray}
&&{\bf u^c}^{--}~,~~ {\bf q}^{-+}~,~~ {\bf e^c}^{--}~,~~~~{\rm and}
\\
&&{\bf \overline{u}^c}^{--}~,~~ {\bf \overline{q}}^{-+}~,
~~ {\bf \overline{e}^c}^{--}~.
\end{eqnarray}
The massless ($n=0$) modes of the gauge scalars $\Phi$, $A_5$s in
${\bf U^c}^{++}_\Sigma$, ${\bf E^c}^{++}_\Sigma$,
and ${\bf \overline{U}^c}^{++}_\Sigma$, ${\bf \overline{E}^c}^{++}_\Sigma$
obtain masses from the gauge coupling
$g^2|\langle \nu_H^c\rangle A_5|^2$, where $\nu_H^c$ ($\nu_H^{c*}$) is
the scalar component of ${\bf \nu^c}^{++}$, ${\bf \overline{\nu}^c}^{++}$.
The gauge bosons in ${\bf Q'}_V^{+-}$
and ${\bf \overline{Q'}}_V^{+-}$ absorb the $A_5$'s from
\begin{eqnarray}
{\bf Q'}_\Sigma^{-+}~, ~~
{\bf \overline{Q'}}_\Sigma^{-+}~.
\end{eqnarray}
We note that the gauge bosons absorb $A_5$'s
carrying the same quantum numbers but opposite parities,
whereas they absorb the Higgs fields with the same parities.
This can be understood from the Lagrangian after symmetry breaking,
${\cal L}\supset (\partial_5A_\mu-\partial_\mu A_5)^2\sim
m_{KK}^2(A_\mu-\frac{1}{m_{KK}}\partial_\mu A_5)^2$ and
${\cal L}\supset g^2v^2(A_\mu-\frac{1}{gv}\partial_\mu a)^2$,
where $m_{KK}$ indicates the KK mass and $a$ is the Goldstone boson of
the scalar Higgs $\phi=(v+\rho)e^{ia/v}/\sqrt{2}$.

Finally, in order to realize the MSSM field contents at low energies,
we should ensure that the three vector-like pairs of Higgs fields,
${\bf u}^{++}$, ${\bf \overline{u}}^{++}$,
${\bf d^c}^{++}$, ${\bf \overline{d}^c}^{++}$, and
${\bf e}^{++}$, ${\bf \overline{e}}^{++}$ are heavy.
This is possible, for example, by introducing at B1
additional brane chiral superfields ${\bf 10}_{-1}^b$,
${\bf \overline{10}}_1^b$, and ${\bf \overline{5}}_{-3}^b$, ${\bf 5}_3^b$
with unit $U(1)_R$ charges.
(Gauge symmetry forbids their couplings to the chiral multiplets
from the 4D $N=2$ vector multiplet.)
%
%

\subsection{$SU(5)\times U(1)_X$~--~$SU(4)_c\times SU(2)_L\times SU(2)_R$}

In this subsection, we take $Z_2\times Z_2'$ elements
to be the $P_2$ and $P_4$.
As already explained, by a $P_2$ operation,
the $SU(5)\times U(1)_X$ generators are assigned positive
parities and their associated gauge multiplets survive at B1.
On the other hand, the $SO(10)$ generators with even parity under $P_4$ are
\begin{eqnarray} \label{so6}
{\bf (8,1)}_{0}^{++}~,~~ {\bf (\overline{3},1)}_{-2/3}^{-+}~,~~
{\bf (3,1)}_{2/3}^{-+}~,~~{\bf (1,1)}_{0}^{++}~; \\
{\bf (1,3)}_{0}^{++}~,~~ {\bf (1,1)}_{1}^{-+}~,~~ {\bf (1,1)}_{-1}^{-+}~,~~
{\bf (1,1)}_{0}^{++}~,   \label{so4}
\end{eqnarray}
all of which survive at B2.  Here the superscripts denote $P_2$ and $P_4$
eigenvalues.
The generators in Eqs.~(\ref{so6}) and (\ref{so4})
correspond to $SO(6)$ and $SO(4)$, respectively.
To see this explicitly, we transform the $SO(10)$ generator
in Eq.~(\ref{so10/z2z2}) with the unitary matrix,
\begin{eqnarray}
U_4=\left(\begin{array}{c|cc|c}
I_{3\times 3} & 0 & 0 & 0 \\ \hline
0 & 0 & I_{2\times 2} & 0 \\
0 & I_{3\times 3} & 0 & 0 \\ \hline
0 & 0 & 0 & I_{2\times 2}
\end{array}\right)_{10\times 10} ~.
\end{eqnarray}
The entries with even parities under $P_4$ are then block-diagonalized,
\begin{eqnarray} \label{422generator}
\left[\begin{array}{cc|cc}
{\bf (8,1)}_{0}^{++} &
{\bf (\overline{3},1)}_{-2/3}^{-+} &
{\bf (3,\overline{2})}_{-5/6}^{+-} & {\bf (3,2)}_{1/6}^{--} \\
{\bf (3,1)}_{2/3}^{-+} &
{\bf (8,1)}_{0}^{++} &
{\bf (\overline{3},\overline{2})}_{-1/6}^{--} &
{\bf (\overline{3},2)}_{5/6}^{+-} \\
\hline
{\bf (\overline{3},2)}_{5/6}^{+-} & {\bf (3,2)}_{1/6}^{--} &
{\bf (1,3)}_{0}^{++}
& {\bf (1,1)}_{1}^{-+} \\
{\bf (\overline{3},\overline{2})}_{-1/6}^{--} &
{\bf (3,\overline{2})}_{-5/6}^{+-} &
{\bf (1,1)}_{-1}^{-+} &
{\bf (1,3)}_{0}^{++}
\end{array}\right]_{10\times 10} ~,
\end{eqnarray}
where we have omitted the two $U(1)$ generators (${\bf (1,1)_0^{++}}$s)
from the diagonal parts.
Using Eq.~(\ref{general}), one can readily check that
the two block-diagonal parts are
$SO(6)\times SO(4)$ ($\sim SU(4)_c\times SU(2)_L\times SU(2)_R$) generators.
The two off diagonal parts in Eq.~(\ref{422generator}) are identified
with each other, and they compose the ${\bf (6,2,2)}$ representations
under $SU(4)_L\times SU(2)_L\times SU(2)_R$.
We conclude that by employing $P_2$ and $P_4$, $SU(5)\times U(1)_X$ and
$SU(4)_L\times SU(2)_L\times SU(2)_R$ are preserved at B1 and B2,
respectively.  The parities of $N=1$ gauge multiplets follow those of
the corresponding generators.

With opposite parities assigned to the chiral multiplets,
the non-vanishing components at B1 are
\begin{eqnarray}
{\bf 10}_\Sigma
&=&{\bf U^c}_\Sigma^{+-}
+{\bf E^c}_\Sigma^{+-}
+{\bf Q}_\Sigma^{++}~,~~{\rm  and}  \\
{\bf \overline{10}}_\Sigma
&=&{\bf \overline{U}^c}_\Sigma^{+-}
+{\bf \overline{E}^c}_\Sigma^{+-}
+{\bf \overline{Q}}_\Sigma^{++} ~,
\end{eqnarray}
while, on B2 brane, the surviving chiral multiplet is
\begin{eqnarray}
{\bf (6,2,2)}_\Sigma
={\bf Q}_\Sigma^{++}
+{\bf \overline{Q}}_\Sigma^{++}
+{\bf Q'}_\Sigma^{-+}
+{\bf \overline{Q'}}_\Sigma^{-+} ~.
\end{eqnarray}
Here we used the notations from Eq.~(\ref{su5}), and
the subscript ``$\Sigma$'' stands for the chiral multiplet.
We show in Table VI the surviving vector and chiral multiplets
on each brane.
\begin{center}
\begin{tabular}{|c|c||c|} \hline
Vector (B1) ~&~ ${\bf 24}_V$, ${\bf 1}_V$ ~&
${\bf G}_V^{++}$,~  ${\bf W}_V^{++}$,~  ${\bf Y}_V^{++}$;~
${\bf X}_V^{++}$,~  ${\bf Q'}_V^{+-}$,~  ${\bf Q'}_V^{+-}$
\\
Chiral (B1) ~&~ ${\bf 10}_\Sigma$, ${\bf \overline{10}}_\Sigma$ ~&
${\bf U^c}_\Sigma^{+-}$,~  ${\bf E^c}_\Sigma^{+-}$,~
${\bf Q}_\Sigma^{++}$;~  ${\bf \overline{U}^c}_\Sigma^{+-}$,~
${\bf \overline{E}^c}_\Sigma^{+-}$,~  ${\bf \overline{Q}}_\Sigma^{++}$
\\ \hline \hline
Vector (B2) &${\bf 15}_V$, ${\bf 3}_V$, ${\bf 3'}_V$&
${\bf G}_V^{++}$,~${\bf U^c}^{-+}_V$,~${\bf \overline{U}^c}_V^{-+}$,
${\bf X}_V^{++}$;~ ${\bf W}_V^{++}$;~
${\bf E^c}^{-+}$,~${\bf \overline{E}^c}^{-+}$,~${\bf Y}_V^{++}$
\\
Chiral (B2) ~&~ ${\bf (6,2,2)}_\Sigma$ ~&
${\bf Q}_\Sigma^{++}$,~  ${\bf \overline{Q}}_\Sigma^{++}$,~
${\bf Q'}_\Sigma^{-+}$,~  ${\bf \overline{Q'}}_\Sigma^{-+}$
\\ \hline
\end{tabular}
\vskip 0.4cm
{\bf Table VI.~} Surviving superfields on each brane
in the $SO(10)$ gauge multiplet.
\end{center}

As seen from Table VI, the vector-like pair,
${\bf Q}_\Sigma^{++}$ and ${\bf \overline{Q}}_\Sigma^{++}$ must be
removed from the low energy spectrum.
They can become massive through spontaneous symmetry breaking
by the bulk Higgs.
Table VII shows the Higgs hypermultiplets and their quantum numbers.
\vskip 0.4cm
\begin{center}
\begin{tabular}{|c||c|c|} \hline
Hypermultiplets & $Z_2\times Z_2'$~ parities & $U(1)_R$
\\ \hline \hline
${\bf 16}_H$ & ~${\bf u^c}^{-+},~{\bf e^c}^{-+},~{\bf q}^{--}~~;~~
{\bf d^c}^{++},~{\bf l}^{+-}~~;~~{\bf \nu^c}^{++} $ & $0$
\\
${\bf 16^c}_H$ & ${\bf u}^{+-}~,~{\bf e}^{+-}~,~{\bf q^c}^{++}~;~~
{\bf d}^{--}~,~{\bf l^c}^{-+}~;~~{\bf \nu}^{--}$ & $0$
\\ \hline
${\bf \overline{16}}_H$ & ~${\bf \overline{u}^c}^{-+},~
{\bf \overline{e}^c}^{-+},~{\bf \overline{q}}^{--}~~;~~
{\bf \overline{d}^c}^{++},~{\bf \overline{l}}^{+-}~~;~~
{\bf \overline{\nu}^c}^{++}$ & $0$
\\
${\bf \overline{16}^c}_H$ & ${\bf \overline{u}}^{+-}~,~
{\bf \overline{e}}^{+-}~,~{\bf \overline{q}^c}^{++}~;~~
{\bf \overline{d}}^{--}~,~
{\bf \overline{l}^c}^{-+}~;~~{\bf \overline{\nu}}^{--}$ & $0$
\\ \hline
\end{tabular}
\vskip 0.4cm
{\bf Table VII.~} $Z_2\times Z_2'$ parities of the bulk Higgs hypermultiplets.
\end{center}
Analogous to the previous case with $SU(5)\times U(1)_X-SU(5)'\times U(1)_X'$,
at B1 they compose $SU(5)\times U(1)_X$ multiplets,
${\bf 10}_{1}$, ${\bf \overline{5}}_{-3}$, ${\bf 1}_{5}$, etc.
At B2 they compose $SU(4)_c\times SU(2)_L\times SU(2)_R$
multiplets such as ${\bf (4,2,1)}$ and ${\bf (\overline{4},1,2)}$
as shown in Table VIII.
\vskip 0.4cm
\begin{center}
\begin{tabular}{|c|c|c|c|} \hline
${\bf 5}_H$, ${\bf 1}_H$ (B1) &
${\bf \overline{5}}_H$, ${\bf \overline{1}}_H$ (B1) & ${\bf 10^c}_H$ (B1) &
${\bf \overline{10}^c}_H$ (B1)
\\ \hline
${\bf d^c}^{++}$, ${\bf l}^{+-}$, ${\bf \nu^c}^{++}$ &
${\bf \overline{d}^c}^{++}$, ${\bf \overline{l}}^{+-}$,
${\bf \overline{\nu}^c}^{++}$ &
${\bf u}^{+-}$, ${\bf e}^{+-}$, ${\bf q^c}^{++}$ &
${\bf \overline{u}}^{+-}$, ${\bf \overline{e}}^{+-}$,
${\bf \overline{q}^c}^{++}$
\\ \hline \hline
${\bf (\overline{4},1,2)}_H$ (B2) & ${\bf (4,1,2)}_H$ (B2) &
${\bf (4^c,2,1)}_H$ (B2) & ${\bf (\overline{4}^c,2,1)}_H$ (B2)
\\ \hline
${\bf u^c}^{-+}$, ${\bf e^c}^{-+}$, ${\bf d^c}^{++}$, ${\bf \nu^c}^{++}$&
${\bf \overline{u}^c}^{-+}$, ${\bf \overline{e}^c}^{-+}$,
${\bf \overline{d}^c}^{++}$, ${\bf \overline{\nu}^c}^{++}$&
${\bf q^c}^{++}$, ${\bf l^c}^{-+}$ &
${\bf \overline{q}^c}^{++}$, ${\bf \overline{l}^c}^{-+}$
\\ \hline
\end{tabular}
\vskip 0.4cm
{\bf Table VIII.~} Surviving Higgs superfields on the branes B1 and B2.
\end{center}

On the two branes, the Higgs superpotentials are
\begin{eqnarray}\label{sp3}
W_{B1}&=&\kappa_1S\bigg({\bf 16}_H{\bf \overline{16}}_H-M_1^2\bigg) ~,
\\ \label{sp4}
W_{B2}&=&\kappa_2S\bigg({\bf 16^c}_H{\bf \overline{16}^c}_H+{\bf 1}{\bf 1'}
-M_2^2\bigg) ~,
\end{eqnarray}
where we schematically wrote the vector-like couplings
of the Higgs multiplets on the two branes,
${\bf 10^c}_H{\bf \overline{10}^c}_H
+{\bf \overline{5}}_H{\bf 5}_H+{\bf 1}_H{\bf \overline{1}}_H$
and ${\bf (\overline{4},1,2)}_H{\bf (4,1,2)}_H
+{\bf (4^c,2,1)}_H{\bf (\overline{4}^c,2,1)}_H$ with arbitrary coefficients
as ${\bf 16}_H{\bf \overline{16}}_H$ and
${\bf 16^c}_H{\bf \overline{16}^c}_H$, respectively.
The gauge singlet superfields ${\bf 1}$, ${\bf 1'}$ are introduced
for the same reason as in section 3.
As in the previous case, the VEVs of ${\bf \nu^c}^{++}$,
${\bf \overline{\nu}^c}^{++}$ lead to the MSSM gauge symmetry, and
generate mass terms of ${\bf X}_V^{++}$, ${\bf Q}_\Sigma^{++}$, and
${\bf \overline{Q'}}_\Sigma^{++}$.
Additional B1 brane superfields ${\bf 10}_1^b$, ${\bf \overline{10}}_{-1}^b$,
and ${\bf \overline{5}}_{-3}^b$, ${\bf 5}_{3}^b$ with $U(1)_R$ charges of
unity, and their bilinear couplings with the Higgs fields at B1
could simply make ${\bf d^c}^{++}$, ${\bf \overline{d}^c}^{++}$,
${\bf q^c}^{++}$, ${\bf \overline{q}^c}^{++}$, etc. heavy.
%
%

Finally another scenario one could consider is one with $SU(5)'\times U(1)_X'$
and $SU(4)_c\times SU(2)_L\times SU(2)_R$ at B1 and B2 respectively.
We will not pursue this any further here.

\section{5D Cosmology}

We consider 5D space-time $x^M=(x^{\mu},y)$, $\mu=0,1,2,3$,
compactified on an $S^1/Z_2$ orbifold, and
the (SUGRA) action is given by
\begin{eqnarray} \label{action}
S=\int d^4x \int_{-y_c}^{y_c}dy~e\bigg[\frac{M_5^3}{2}~R_5+{\cal L}_B
+\sum_{i=I,II}\frac{\delta(y-y_i)}{e_5^5}\bigg(\frac{M_i^2}{2}~\bar{R}_4
+{\cal L}_i\bigg)
\bigg] ~,
\end{eqnarray}
where $R_5$ ($\bar{R}_4$) stands for the 5 dimensional (4 dimensional)
Einstein-Hilbert term, ${\cal L}_B$ (${\cal L}_I$, ${\cal L}_{II}$)
denotes some unspecified bulk (brane)
contributions to the full Lagrangian, and $y_I=0$, $y_{II}=y_c$ indicate
the brane positions.
The brane scalar curvature term $\bar{R}_4(\bar{g}_{\mu\nu})$
is defined through the induced metric,
$\bar{g}_{\mu\nu}(x)\equiv g_{\mu\nu}(x,y=0)$ ($\mu,\nu=0,1,2,3$).
The brane-localized Einstein-Hilbert terms\footnote{
The importance of the
brane-localized 4D Einstein-Hilbert term, especially for generating 4D gravity
in a higher dimensional non-compact flat space was first noted
in Ref.~\cite{braneR}.
} in Eq.~(\ref{action})
are allowed also in SUGRA, but should, of course, be accompanied
by brane gravitino kinetic terms as well as other terms,
as spelled out in the off-shell SUGRA formalism \cite{kyae}.
Here we assume that the bulk cosmological constant is zero.

For the cosmological solution let us take the following metric ansatz,
\begin{eqnarray} \label{metric}
ds^2=\beta^2(t,y)\bigg(-dt^2+a^2(t)~d\vec{x}^2\bigg)+dy^2 ~,
\end{eqnarray}
which shows that the three dimensional space is homogeneous and isotropic.
The non-vanishing components of the 5D Einstein tensor derived
from Eq.~(\ref{action}) are
\begin{eqnarray}
G^0_0&=&3\bigg[\bigg(\frac{\beta''}{\beta}\bigg)
+\bigg(\frac{\beta'}{\beta}\bigg)^2~\bigg]
-\frac{3}{\beta^2}\bigg[\bigg(\frac{\dot{\beta}}{\beta}
+\frac{\dot{a}}{a}\bigg)^2~\bigg] \nonumber \\
&&-\sum_{i=I,II}\delta(y-y_i)\frac{M_i^2}{M_5^3}\frac{3}{\beta^2}\bigg[
\bigg(\frac{\dot{\beta}}{\beta}
+\frac{\dot{a}}{a}\bigg)^2~\bigg]
~, \label{eins00} \\
G^i_i&=&3\bigg[\bigg(\frac{\beta''}{\beta}\bigg)
+\bigg(\frac{\beta'}{\beta}\bigg)^2~\bigg]
-\frac{1}{\beta^2}\bigg[2\frac{\ddot{\beta}}{\beta}+2\frac{\ddot{a}}{a}
+4\frac{\dot{\beta}}{\beta}\frac{\dot{a}}{a}
-\bigg(\frac{\dot{\beta}}{\beta}\bigg)^2+\bigg(\frac{\dot{a}}{a}\bigg)^2
~\bigg] \nonumber \\
&&-\sum_{i=I,II}\delta(y-y_i)\frac{M_i^2}{M_5^3}\frac{1}{\beta^2}\bigg[
2\frac{\ddot{\beta}}{\beta}+2\frac{\ddot{a}}{a}
+4\frac{\dot{\beta}}{\beta}\frac{\dot{a}}{a}
-\bigg(\frac{\dot{\beta}}{\beta}\bigg)^2+\bigg(\frac{\dot{a}}{a}\bigg)^2
~\bigg]
~, \label{einsii} \\
G^5_5&=&6\bigg[\bigg(\frac{\beta'}{\beta}\bigg)^2~\bigg]-\frac{3}{\beta^2}
\bigg[\frac{\ddot{\beta}}{\beta}+\frac{\ddot{a}}{a}
+3\frac{\dot{\beta}}{\beta}\frac{\dot{a}}{a}+\bigg(\frac{\dot{a}}{a}\bigg)^2
~\bigg]
~, \label{eins55} \\
G_{05}&=&-3\bigg[\bigg(\frac{\beta'}{\beta}\bigg)^{\cdot}~ \bigg] ~,
\label{eins05}
\end{eqnarray}
where primes and dots respectively denote derivatives with respect to
$y$ and $t$, and the terms accompanied by delta functions
arise from the brane-localized Einstein-Hilbert terms.

Let us first discuss inflation under this set up.
Since 5D $N=1$ SUSY does not allow a superpotential
(and the corresponding F-term scalar potential) in the bulk,
we introduce the inflaton scalar potentials $V_{I,II}(\phi)$ ($\geq 0$)
on the two branes where only 4D $N=1$ SUSY is preserved~\cite{KS,king}.
The energy-momentum tensor during inflation is given by
\begin{eqnarray}
T^0_0~=~T^i_i&=&-\delta(y)\frac{V_{I}}{M_5^3}
-\delta(y-y_c)\frac{V_{II}}{M_5^3} ~,   \\
T^5_5&=&0 ~,
\end{eqnarray}
where $i=1,2,3$, and $V_{I}$, $V_{II}$ are the scalar potentials on B1 and B2
that are suitably chosen
to provide a large enough number of e-foldings to resolve the horizon and
flatness problems.
The end of inflation is marked by the breaking of the `slow roll' conditions,
and the inflaton rolls quickly to the true suepersymmetric vacuum
with flat 4D space-time.
Thus, for the inflationary epoch
it is sufficient to consider only scalar potentials
in the energy-momentum tensor.
We will discuss more general cases later.
%
%

The exact inflationary solution is~\cite{KS}
\begin{eqnarray}
&&\beta(y)=H_0|y|+c ~, \label{beta} \\
&&a(t)=e^{H_0t} ~, \label{a}
\end{eqnarray}
where $H_0$ ($>0$) is the Hubble constant during inflation.
%
%
The integration constant $c$ in Eq.~(\ref{beta}) can be normalized
to unity without loss of generality.
%
%
%
The introduction of the brane-localized Einstein-Hilbert terms
do not affect the bulk solutions Eqs.~(\ref{beta}) and (\ref{a}),
but they modify the boundary conditions.
The solution $\beta(y)$ should satisfy the following boundary conditions
at $y=0$ and $y=y_c$,
\begin{eqnarray} \label{bdy1}
\frac{V_{I}}{6M_5^3}&=&-H_0+\frac{1}{2}\frac{M_I^2}{M_5^3}~H_0^2~, \\
\frac{V_{II}}{6M_5^3}&=&\frac{H_0}{1+H_0y_c}
+\frac{1}{2}\frac{M_{II}^2}{M_5^3}~\frac{H_0^2}{(1+H_0y_c)^2}~.
\label{bdy2}
\end{eqnarray}
Thus, $H_0$ and $y_c$ are determined by $V_I$ and $V_{II}$.
Note that the brane cosmological constants (scalar potentials) $V_I$
and $V_{II}$ are related to the Hubble constant $H_0$.
While the non-zero brane cosmological constants are responsible for
inflating the 3-space, their subsequent vanishing
restores SUSY and guarantees the flat 4D space-time.
Since $V_{II}$ must be zero when $V_I=0$,
it is natural that the scalar field controlling the end of inflation is
introduced in the bulk.
With SUSY broken at low energies, the minima of the inflaton potentials
on both branes should be fine-tuned to zero~\cite{selftun}.

From Eqs.~(\ref{bdy1})--(\ref{bdy2}) we note that
in the absence of the brane-localized Einstein-Hilbert term at $y=0$,
the inflaton potentials (brane cosmological constants)
$V_I$ and $V_{II}$ should have opposite signs.
However, a suitably large value of $M_I/M_5$~\cite{braneR}
can flip the sign of $V_I$~\cite{KS}, so that both $V_I$ and
$V_{II}$ are positive.
%
%
Thus, a brane-localized Einstein-Hilbert term at $y=0$ seems essential
for successful F-term inflation in the 5D SUSY framework.
Its introduction does not conflict with any symmetry,
and in~\cite{KS} a simple model for realizing a large ratio $M_I/M_5$
was proposed.

The 4D reduced Planck mass ($\equiv (M_{\rm Planck}/8\pi)^{1/2}$)
is given by
%
\begin{eqnarray}
M_{P}^2&=& M_5^3\int_{-y_c}^{y_c}dy\beta^2+M_I^2+M_{II}^2  \nonumber \\
&=&M_5^3y_c\bigg(\frac{2}{3}H_0^2y_c^2+2H_0y_c+2\bigg)+M_I^2+M_{II}^2 ~,
\end{eqnarray}
while the 4D effective cosmological constant is calculated to be
\begin{eqnarray}
\Lambda_{\rm eff}&=&\int_{-y_c}^{y_c}dy\beta^4\bigg[M_5^3\bigg(
4\bigg(\frac{\beta''}{\beta}\bigg)+6\bigg(\frac{\beta'}{\beta}\bigg)^2\bigg)
+\delta(y)V_I+\delta(y-y_c)V_{II}\bigg]  \nonumber \\
&=&3H_0^2\bigg[M_5^3y_c\bigg(\frac{2}{3}H_0^2y_c^2
+2H_0y_c+2\bigg)+M_I^2+M_{II}^2\bigg] \nonumber \\
&=&3H_0^2M_P^2 ~,
\end{eqnarray}
which vanishes when $V_I=V_{II}=0$.

After inflation, the inflaton decays into brane
and (subsequently) bulk fields,
which reheat the whole 5 dimensional universe.
To quantify the inflaton and radiation (or matter) dominated epochs,
we use the fluid approximation,
\begin{eqnarray} \label{fluid}
T^{M}\,_N=\frac{1}{M_5^3}
\left(\begin{array}{cccc|c}
-\rho & 0 & 0 & 0 & T^0\,_{5}\\
0 & p & 0 & 0 & 0 \\
0 & 0 & p & 0 & 0 \\
0 & 0 & 0 & p & 0 \\ \hline
T^5\,_{0} & 0 & 0 & 0 & P_5
\end{array}\right) ~,
\end{eqnarray}
%
%
where $\rho$ and $p$ are contributed by bulk and brane matter,
\begin{eqnarray} \label{matter}
\rho&\equiv&\frac{1}{2y_c}\rho_B+\sum_{i=I,II}\delta(y-y_i)\rho_{i} ~,\\
p&\equiv&\frac{1}{2y_c}p_B+2y_c\sum_{i=I,II}\delta(y-y_i)p_i ~.
\label{matter2}
\end{eqnarray}
Note that in Eq.~(\ref{fluid}) the non-zero off-diagonal components,
$T^0\,_5$ ($=\frac{-1}{\beta^2}T_{05}$) and $T^5\,_0$ ($=T_{05}$)
are considered.
In Eqs.~(\ref{matter}) and (\ref{matter2}),
we normalize $\rho_B$ and $p_B$ with the circumference of
the extra dimension, so their components have the same mass dimension
as their brane counterparts.
%
%
%
%
With Eqs.~(\ref{eins00})--(\ref{eins05}) and (\ref{fluid}),
the 5D ``Friedmann-like'' equations are readily written,
\begin{eqnarray}
\frac{1}{2y_cM_5^3}~\rho_B&=&\frac{3}{\beta^2}\bigg[
\bigg(\frac{\dot{\beta}}{\beta}\bigg)^2+2\frac{\dot{\beta}}{\beta}H
+\bigg(H^2-h^2\bigg)\bigg] ~,  \label{rhoB}  \\
\frac{1}{M_5^3}~\rho_{I}&=&\bigg[3\frac{M_I^2}{M_5^3}H^2-6h\bigg] ~, \\
\frac{1}{M_5^3}~\rho_{II}&=&\bigg[\frac{M_{II}^2}{M_5^3}
\frac{3}{\beta^2}\bigg(\bigg(\frac{\dot{\beta}}{\beta}\bigg)^2
+2\frac{\dot{\beta}}{\beta}H+H^2\bigg)+6\frac{h}{\beta}\bigg]_{y=y_c} ~,  \\
\frac{1}{2y_cM_5^3}~p_B&=&-\frac{1}{\beta^2}\bigg[
2\frac{\ddot{\beta}}{\beta}-\bigg(\frac{\dot{\beta}}{\beta}\bigg)^2
+4\frac{\dot{\beta}}{\beta}H+2\dot{H}+3\bigg(H^2-h^2\bigg)\bigg] ~, \\
\frac{1}{M_5^3}~p_{I}&=&-\bigg[\frac{M_I^2}{M_5^3}\bigg(
2\dot{H}+3H^2\bigg)-6h\bigg]  ~, \\
\frac{1}{M_5^3}~p_{II}&=&-\bigg[\frac{M_{II}^2}{M_5^3}\frac{1}{\beta^2}\bigg(
2\frac{\ddot{\beta}}{\beta}-\bigg(\frac{\dot{\beta}}{\beta}\bigg)^2
+4\frac{\dot{\beta}}{\beta}H
+2\dot{H}+3H^2\bigg)+6\frac{h}{\beta}\bigg]_{y=y_c} ~, \\
\frac{1}{2y_cM_5^3}~P_5&=&-\frac{3}{\beta^2}\bigg[
\frac{\ddot{\beta}}{\beta}+3\frac{\dot{\beta}}{\beta}H+\dot{H}
+2\bigg(H^2-h^2\bigg)\bigg] ~, \\
\frac{1}{2y_cM_5^3}~T_{05}&=&-3~{\rm sgn}(y)\bigg[\frac{\dot{h}}{\beta}
-\frac{h}{\beta}\frac{\dot{\beta}}{\beta}\bigg] ~. \label{T05}
\end{eqnarray}
Here ${\rm sgn}(y)\equiv 1(-1)$ for $y>0(<0)$, and
\begin{eqnarray}
H(t)&\equiv& \frac{\dot{a}}{a} ~, \\
\beta(t,y)&=&h(t)|y|+1 ~.
\end{eqnarray}
For $M_I>M_5, M_{II}$, and $H>>h$,
the brane matter contribution from B1 is dominant, and
Eqs.~(\ref{rhoB})--(\ref{T05}) reduce to
the approximate 4D Friedmann equations,
\begin{eqnarray} \label{fried1}
&&\bigg(\frac{\dot{a}}{a}\bigg)^2\approx\frac{1}{3M_4^2}~\rho_I~, \\
&&~~\frac{\ddot{a}}{a}\approx\frac{-1}{6M_4^2}\bigg(\rho_I+3p_I
\bigg)~. \label{fried2}
\end{eqnarray}
Eqs.~(\ref{rhoB})--(\ref{T05}) satisfy
the energy-momentum conservation law,
$\nabla_MT^M_N=0$ whose $N=0$ and $N=5$ components are~\cite{kim-kyae}
\begin{eqnarray} \label{conserv1}
\dot{\rho}+3\bigg(\frac{\dot{\beta}}{\beta}+H\bigg)(\rho+p)&=&
T^{5'}\,_0+4\frac{\beta'}{\beta}T^{5}\,_0
\nonumber \\
&=&2y_cM_5^3\bigg[G_{05}'+4\frac{\beta'}{\beta}G_{05}\bigg] ~,
\\ \label{conserv2}
P_5'+\frac{\beta'}{\beta}\bigg(4P_5-3p+\rho\bigg)&=&
-\dot{T}^{0}\,_5-\bigg(4\frac{\dot{\beta}}{\beta}+3H\bigg)T^{0}\,_5
\nonumber \\
&=&2y_cM_5^3\frac{1}{\beta^2}\bigg[\dot{G}_{05}+\bigg(
2\frac{\dot{\beta}}{\beta}+3H\bigg)G_{05}\bigg] ~.
\end{eqnarray}
The inflaton contributes to the energy momentum tensor Eq.~(\ref{fluid}),
\begin{eqnarray} \label{EM}
T_{MN}\equiv ~T^{\rm inf}_{MN}+T^{\rm m}_{MN} ~,
\end{eqnarray}
where $T^{\rm inf}_{MN}$ denotes the contributions to
the energy momentum tensor from the inflaton $\phi(t,y)$,
\begin{eqnarray} \label{inflaton}
T^{\rm inf}_{MN}&\equiv& \frac{1}{2y_c}\partial_M\phi\partial_N\phi
-\frac{1}{4y_c}g_{MN}\partial^P\phi\partial_P\phi  \\
&&+\sum_{i=I,II}\delta(y-y_i)\delta_M^\mu\delta_N^\nu\bigg[
\partial_{\mu}\phi\partial_\nu\phi
-g_{\mu\nu}\bigg(\frac{1}{2}\partial^\lambda\phi\partial_\lambda\phi
+V_i(\phi)\bigg)\bigg]~,  \nonumber
\end{eqnarray}
and $T^{\rm m}_{MN}$ is assumed to have the same form as Eq.~(\ref{fluid}).
%
%
The conservation law $\nabla^MT^{\rm inf}_{MN}=0$
gives rise to the scalar field equation
in the presence of both the brane and bulk kinetic terms.
If only the inflaton potentials on the branes, $V_I(\phi)$ and $V_{II}(\phi)$
are dominant in Eq.~(\ref{EM}),  one can check that
the solutions reduce to Eqs.~(\ref{beta}) and (\ref{a}), namely,
$H=h={\rm constant}~(=H_0)$.
The inflaton decay produces $T^{\rm m}_{MN}$.

%
%

We have tacitly assumed that the interval separating the two branes
(orbifold fixed points) remains fixed during inflation.
The dynamics of the orbifold fixed points,
unlike the D-brane case~\cite{Dbrane},
is governed only by the $g_{55}(x,y)$ component of the metric tensor.
The real fields $e_5^5$, $B_5$, and the chiral fermion $\psi^2_{5R}$
in 5D gravity multiplet are assigned even parity under $Z_2$ \cite{kyae},
and they compose an $N=1$ chiral multiplet on the branes.
The associated superfield can acquire a superheavy mass and
its scalar component can develop a vacuum expectation value (VEV)
on the brane.
With superheavy brane-localized mass terms,
the low-lying Kaluza-Klein (KK) mass spectrum is shifted
so that even the lightest mode
obtains a compactification scale mass~\cite{localmass}.
Since this mass is much greater than $H_0$
the interval distance is stable even during inflation.
This stabilization of the interval distance in turn leads to
the stabilization of the warp factor $\beta(y)$.
This is because the fluctuation $\delta \beta(y)$
of the warp factor near the solution
in Eq.~(\ref{beta}) turns out to be proportional to
the interval length variation $\delta g_{55}$
from the linearized 5D Einstein equation~\cite{chacko}.

So far we have discussed only $S^1/Z_2$ orbifold compactification.
The results can be directly extended to $S^1/(Z_2\times Z_2')$.
Within the framework discussed in this section,
we can accommodate any promising 4D SUSY inflationary model.
We consider one particular model below which comes from compactifying
$SO(10)$ on an $S^1/(Z_2\times Z_2')$.

\section{Inflation and Leptogenesis in 4D SUSY Model}

The 4D inflationary model is best illustrated by considering
the following superpotential which allows the breaking of some gauge
symmetry $G$ down to $SU(3)_c\times SU(2)_L\times U(1)_Y$,
keeping $N=1$ SUSY intact \cite{hybrid, lyth}:
\begin{eqnarray} \label{simplepot}
W_{\rm infl}=\kappa S(\phi\bar{\phi}-M^2) ~.
\end{eqnarray}
Here $\phi$ and $\bar{\phi}$ represent superfields
whose scalar components acquire non-zero vacuum expectation values (VEVs).
For the particular example of $G=H$ above,
they belong to the ${\bf (\overline{4}, 1,2)}$ and ${\bf (4,1,2)}$
representations of $SU(4)_c\times SU(2)_L\times SU(2)_R$.
The $\phi$, $\bar{\phi}$ VEVs break $SU(4)_c\times SU(2)_L\times SU(2)_R$ 
to the MSSM gauge group.
The singlet superfield $S$ provides the scalar field that drives inflation.
Note that by invoking a suitable $R$ symmetry $U(1)_R$,
the form of $W$ is unique at the renormalizable level, and it
is gratifying to realize that $R$ symmetries naturally occur
in (higher dimensional) SUSY theories
and can be appropriately exploited.
From $W$, it is straightforward to show that the SUSY minimum
corresponds to non-zero (and equal in magnitude) VEVs
for $\phi$ and $\bar{\phi}$, while $\langle S\rangle =0$~\cite{king}.
(After SUSY breaking {\it $\grave{a}$ la} $N=1$ supergravity
(SUGRA), $\langle S\rangle$ acquires a VEV
of order $m_{3/2}$ (gravitino mass)).

An inflationary scenario is realized in the early universe
with both $\phi$, $\bar{\phi}$ and $S$ displaced
from their present day minima.
Thus, for $S$ values in excess of the symmetry breaking scale $M$,
the fields $\phi$, $\bar{\phi}$ both vanish,
the gauge symmetry is restored, and a potential energy density proportional
to $M^4$ dominates the universe. With SUSY thus broken, there are
radiative corrections from the $\phi$-$\bar{\phi}$ supermultiplets
that provide logarithmic corrections to the potential which drives inflation.
In one loop approximation \cite{hybrid, coleman},
\begin{eqnarray}\label{scalarpot}
V\approx V_{\rm tree}+\kappa^2M^4\frac{\kappa^2N}{32\pi^2}\bigg[
2{\rm ln}\frac{\kappa^2|S|^2}{\Lambda^2}+(z+1)^2{\rm ln}(1+z^{-1})
+(z-1)^2{\rm ln}(1-z^{-1})\bigg]~,
\end{eqnarray}
where $z=x^2=|S|^2/M^2$, $\Lambda$ denotes a renormalization mass scale and
$N$ denotes the dimensionality of the $\phi$, $\bar{\phi}$ representations.
From Eq.~(\ref{scalarpot}) the quadrupole anisotropy is
found to be~\cite{hybrid,hybrid2}
\begin{eqnarray}\label{T}
\bigg(\frac{\delta T}{T}\bigg)_Q\approx \frac{8\pi}{\sqrt{N}}
\bigg(\frac{N_Q}{45}\bigg)^{1/2}\bigg(\frac{M}{M_{\rm Planck}}\bigg)^2
x_Q^{-1}y_Q^{-1}f(x_Q^2)^{-1} ~.
\end{eqnarray}
The subscript $Q$ is there to emphasize the epoch of horizon crossing,
$y_Q\approx x_Q(1-7/12x_Q^2+\cdots)$, $f(x_Q^2)^{-1}\approx 1/x_Q^2$,
for $S_Q$ sufficiently larger than $M$,
and $N_Q\approx 45-60$ denotes the e-foldings needed to resolve the horizon
and flatness problems.
From the expression for $\delta T/T$ in Eq.~(\ref{T}) and comparison with the
COBE result $(\delta T/T)_Q\approx 6.6\times 10^{-6}$~\cite{cobe},
it follows that the gauge symmetry breaking scale $M$ is
close to $10^{16}$ GeV.  Note that $M$ is associated in our $SO(10)$ example
with the breaking scale of $SU(4)_c\times SU(2)_L\times SU(2)_R$ 
(in particular the $B-L$ breaking scale),
which need not exactly coincide with the SUSY GUT scale.
We will be more specific about $M$ later in the text.

The relative flatness of the potential ensures that the primordial density
fluctuations are essentially scale invariant.
Thus, the scalar spectral index $n$ is
0.98 for the simplest example based on $W$ in Eq.~(\ref{simplepot}).
It should be noted that the inclusion of supergravity corrections
can, in some cases, lead to a spectral index larger than unity
[For a recent discusion and additional references see Ref.~\cite{ns}.].

Several comments are in order:

\noindent $\bullet$ The 50-60 e-foldings required to solve the horizon and
flatness problems occur when the inflaton field $S$ is relatively close
(to within a factor of order 1-10) to the GUT scale.
Thus, Planck scale corrections can be safely ignored.

\noindent $\bullet$ For the case of minimal K${\rm\ddot{a}}$hler potential,
the SUGRA corrections do not spoil the scenario, 
but some interesting restrictions on $\kappa$ can be found~\cite{ns}.    
which is a non-trivial result~\cite{hybrid2}.
More often than not, SUSY inflationary scenarios fail to work
in the presence of SUGRA corrections which tend to spoil the flatness
of the potential needed to realize inflation.

\noindent $\bullet$ Turning to the subgroup 
$SU(4)_c\times SU(2)_L\times SU(2)_R$ of $SO(10)$,
one needs to take into account the fact that the spontaneous breaking of 
$SU(4)_c\times SU(2)_L\times SU(2)_R$ 
produces magnetic monopoles that carry two quanta of
Dirac magnetic charge~\cite{magg}.
An overproduction of these monopoles at or near
the end of inflation is easily avoided, say by introducing an additional
(non-renormalizable) term $S(\phi\bar{\phi})^2$ in $W$,
which is permitted by the $U(1)_R$ symmetry.
The presence of this term ensures the absence of monopoles as explained
in Ref.~\cite{khalil}. Note that the monopole problem is also
avoided by choosing a different subgroup of $SO(10)$.
In a separate publication, we will consider a scenario based on the
$SU(3)_c\times SU(2)_L\times U(1)_Y\times U(1)_X$ subgroup of $SO(10)$
whose breaking does not lead to monopoles.
Another interesting candidate is
$SU(3)_c\times SU(2)_L\times SU(2)_R\times U(1)_{B-L}$.
The salient features of the model are not affected by the monopole
problem~\cite{khalil}.

\noindent $\bullet$ At the end of inflation the scalar fields
$\phi$, $\bar{\phi}$, and $S$ oscillate about their respective minima.
Since the $\phi$, $\bar{\phi}$ belong respectively
to the ${\bf (\overline{4},1,2)}$
and ${\bf (4,1,2)}$ of $SU(4)_c\times SU(2)_L\times SU(2)_R$,
they decay exclusively into right handed neutrinos
via the superpotential couplings,
\begin{eqnarray} \label{nu}
W=\frac{\gamma_{i}}{M_P}\bar{\phi}\bar{\phi}F^c_iF^c_i ~,
\end{eqnarray}
where the matter superfields $F^c_i$ belong to
the ${\bf (\overline{4},1,2)}$ representation of 
$SU(4)_c\times SU(2)_L\times SU(2)_R$, and
$M_P\equiv M_{\rm Planck}/\sqrt{8\pi}=2.44\times 10^{18}$ GeV denotes
the reduced Planck mass, and $\gamma_{i}$ are dimensionless coefficients.
We will have more to say about inflaton decay,
the reheat temperature, as well as leptogenesis taking account of the recent
neutrino oscillation data.
However, we first wish to provide a 5D setting for this inflationary scenario.

After inflation is over, the oscillating system consists of the complex scalar
fields $\Phi=(\delta\bar{\phi}+\delta\phi)$,
where $\delta\bar{\phi}=\bar{\phi}-M$ ($\delta \phi=\phi-M$), and $S$,
both with masses equal to $m_{\rm infl}=\sqrt{2}\kappa M$.
Through the superpotential couplings in Eq.~(\ref{nu}),
these fields decay into a pair of right handed neutrinos and sneutrinos
respectively, with an approximate decay width \cite{khalil}
\begin{eqnarray}\label{decay}
\Gamma\sim \frac{m_{\rm infl}}{8\pi}\bigg(\frac{M_i}{M}\bigg)^2 ~,
\end{eqnarray}
where $M_i$ denotes the mass of the heaviest right handed neutrino
with $2M_i<m_{\rm infl}$, so that the inflaton decay is possible.
Assuming an MSSM spectrum below the GUT scale, the reheat temperature
is given by \cite{hybrid3}
\begin{eqnarray} \label{temp}
T_r\approx \frac{1}{3}\sqrt{\Gamma M_P}
\approx\frac{1}{12}\bigg(\frac{55}{N_Q}\bigg)^{1/4}\sqrt{y_Q}M_i ~.
\end{eqnarray}
For $y_Q\sim$ unity (see below), and $T_r\lapproxeq 10^{9.5}$ GeV
from the gravitino constraint~\cite{gravitino,newreheat},
we require $M_i\lapproxeq 10^{10}-10^{10.5}$ GeV.

In order to decide on which $M_i$ is involved in the decay~\cite{pati},
let us start with atmospheric neutrino ($\nu_\mu-\nu_\tau$) oscillations and
assume that the light neutrinos exhibit an hierarchical mass pattern
with $m_3>>m_2>>m_1$.
Then $\sqrt{\Delta m^2_{\rm atm}}\approx m_3\approx m_{D3}^2/M_3$,
where $m_{D3}$ ($=m_t(M)$)
denotes the third family Dirac mass which equals the asymptotic top quark
mass due to $SU(4)_c$.  We also assume a mass hierarchy in the right handed
sector, $M_3>>M_2>>M_1$.
The mass $M_3$ arises from the superpotential coupling Eq.~(\ref{nu})
and is given by
$M_3= 2\gamma_{3}M^2/M_P\sim 10^{14}~~{\rm GeV}$,
for $M\sim 10^{16}$ GeV and $\gamma_{3}\sim$ unity.  This value of $M_3$ is in
the right ball park to generate an $m_3\sim \frac{1}{20}$ eV
($\sim \sqrt{\Delta m_{\rm atm}^2}$)~\cite{atmos},
with $m_t(M)\sim 110$ GeV \cite{hybrid3}.
It follows from (\ref{temp}) that $M_i$ in (\ref{decay}) cannot be identified
with the third family right handed neutrino mass $M_3$.
It should also not correspond to the second family neutrino mass $M_2$
if we make the plausible assumption that the second generation Dirac mass
should lie in the few GeV scale.
The large mixing angle MSW solution of the solar neutrino
problem requires that
$\sqrt{\Delta m_{\rm solar}^2}\approx m_2\sim {\rm GeV}^2/M_2
\sim \frac{1}{160}~{\rm eV}$~\cite{solar},
so that $M_2\gapproxeq 10^{11}-10^{12}$ GeV.
Thus, we are led to conclude~\cite{pati} that
the inflaton decays into the lightest (first family) right handed neutrino
with mass
\begin{eqnarray}\label{M1}
M_1\sim 10^{10}-10^{10.5} ~{\rm GeV} ~,
\end{eqnarray}
such that $2M_1<m_{\rm infl}$.

The constraint $2M_2> m_{\rm infl}$ yields
$y_Q\lapproxeq 3.34 \gamma_{2}$,
where $M_2=2\gamma_{2}M^2/M_P$.  We will not provide here a comprehensive
analysis of the allowed parameter space but will be content to present
a specific example, namely
\begin{eqnarray}\label{value}
M\approx 8\times 10^{15}~{\rm GeV}~,~~\kappa\approx 10^{-3}~,~~
m_{\rm infl}\sim 10^{13}~{\rm GeV}~(\sim M_2)~,
\end{eqnarray}
with $y_Q\approx 0.4$ (corresponding to $x_Q$ near unity, so that the inflaton
$S$ is quite close to $M$ during the last 50--60 e-foldings).

Note that typically $\kappa$ is of order
$10^{-2}$-- few $\times 10^{-4}$~\cite{khalil,ns},
so that the vacuum energy density during inflation is
$\sim 10^{-4}-10^{-8}~M_{\rm GUT}^4$.
Thus, in this class of models the gravitational wave contribution to
the quadrupole anisotropy $(\delta T/T)_Q$ is essentially negligible
($\lapproxeq 10^{-8}$).
With $\kappa\sim{\rm few}\times 10^{-4}$ ($10^{-3}$),
the scalar spectral index $n\approx 0.99$ ($0.98$).

The decay of the (lightest) right handed neutrinos generates
a lepton asymmetry which is given by \cite{lasymm}
\begin{eqnarray} \label{lasymm}
\frac{n_L}{s}\approx \frac{10}{16\pi}\bigg(\frac{T_r}{m_{\rm infl}}\bigg)
\bigg(\frac{M_1}{M_2}\bigg)
\frac{c_{\theta}^2s_{\theta}^2~{\rm sin}2\delta~(m_{D2}^2-m_{D1}^2)^2}
{|\langle h\rangle|^2(m_{D2}^2s_{\theta}^2+m_{D1}^2c_{\theta}^2)} ~,
\end{eqnarray}
where the VEV $|\langle h\rangle|\approx 174$ GeV (for large ${\rm tan}\beta$),
$m_{D1,2}$ are the neutrino Dirac masses (in a basis in which they are
diagonal and positive), and $c_\theta\equiv {\rm cos}\theta$,
$s_\theta\equiv {\rm sin}\theta$, with $\theta$ and $\delta$ being the rotation
angle and phase which diagonalize the Majorana mass matrix of
the right handed neutrinos.
Assuming $c_\theta$ and $s_\theta$ of comparable magnitude,
taking $m_{D2}>>m_{D1}$, and using (\ref{M1}) and (\ref{value}),
Eq.~(\ref{lasymm}) reduces to
\begin{eqnarray}
\frac{n_L}{s}\approx 10^{-8.5}c_\theta^2 ~{\rm sin}2\delta ~
\bigg(\frac{T_r}{10^{9.5}~{\rm GeV}}\bigg)
\bigg(\frac{M_1}{2\cdot 10^{10.5}~{\rm GeV}}\bigg)
\bigg(\frac{10^{13}~{\rm GeV}}{M_2}\bigg)
\bigg(\frac{m_{D2}}{10~{\rm GeV}}\bigg)^2~,
\end{eqnarray}
which can be in the correct ball park to account for the observed baryon
asymmetry $n_B/s$ ($\approx -28/79 ~n_L/s$)~\cite{sep}.

\section{F-term Inflation in 5D $SO(10)$} 

In this section, we present a realistic 5D $SO(10)$ model 
in which the inflationary scenario described by the superpotential $W$
in Eq.~(\ref{simplepot}) can be realized.
We assume compactification on an orbifold $S^1/(Z_2\times Z_2')$,
such that on the two fixed points (branes) we have the gauge symmetries
$SO(10)$ and $SU(4)_c\times SU(2)_L\times SU(2)_R$ respectively.
To realize the MSSM gauge group at low energies,
we introduce the Higgs hypermultiplets ${\bf 16}_H$ and
${\bf \overline{16}}_H$ in the bulk with $Z_2\times Z_2'$ parities,
\begin{eqnarray}\label{1}
&&{\bf 16}_H~={\bf (\overline{4},1,2)}_H^{++}~+~{\bf (4,2,1)}_H^{+-} ~~,\\
&&{\bf 16^c}_H={\bf (\overline{4},1,2)}_H^{c--}+~{\bf (4,2,1)}_H^{c-+} ~,\\
&&{\bf \overline{16}}_H~   \label{3}
={\bf(4,1,2)}_H^{++}~+~{\bf (\overline{4},2,1)}_H^{+-}~~, \\
&&{\bf \overline{16}^c}_H
={\bf(4,1,2)}_H^{c--}+~{\bf (\overline{4},2,1)}_H^{c-+} ~.
\end{eqnarray}
The relevant superpotentials on the two branes, B1 ($SO(10)$ brane) and
B2 ($SU(4)_c\times SU(2)_L\times SU(2)_R$ brane) are:
\begin{eqnarray}
&&W_{B1}=\kappa S\bigg({\bf 16}_H{\bf \overline{16}}_H-M_1^2\bigg) ~, \\
&&W_{B2}=\kappa S\bigg(c_1H^c\overline{H}^c
+c_2{\bf 1}{\bf 1'}
-M_2^2\bigg)+c_3\Sigma H^c\overline{H}^c ~,  \label{wb2}
\end{eqnarray}
where $H^c\equiv {\bf (\overline{4},1,2)}_H^{++}$,
$\overline{H}^c\equiv {\bf(4,1,2)}_H^{++}$, and $c_1$, $c_2$, $c_3$ are
dimensionless couplings.
In $W_{B2}$, we exhibit only the chiral multiplets with $(++)$ parities
of ${\bf 16}_H$, ${\bf \overline{16}}_H$ which contain massless modes,
since the heavy KK modes would be decoupled.
Since the inflaton $S$ is a bulk superfield,
it participates in both superpotentials.
In Eq.~(\ref{wb2}), a pair of singlet superfields ${\bf 1}$, ${\bf 1'}$ and
a superfield $\Sigma$ in the adjoint representation ${\bf (15,1,1)}$
with suitable $U(1)_R$ charges are introduced on B2.

During inflation, $S$ and $\Sigma$ develop VEVs
($\langle S\rangle>M_1,M_2$),
while $\langle{\bf 16}_H\rangle=\langle{\bf \overline{16}}_H\rangle
=\langle H^c\rangle=\langle\overline{H}^c\rangle
=\langle{\bf 1}\rangle=\langle{\bf 1'}\rangle=0$.
As shown in Refs.~\cite{KS,ks2},
positive vacuum energies localized on the branes could trigger
exponential expansion of the three space,
in the presence of a brane-localized Einstein-Hilbert term.
Due to a non-zero VEV of $\Sigma$ during inflation, the $SU(4)_c$ factor
in $SU(4)_c\times SU(2)_L\times SU(2)_R$ is spontaneously broken to
$SU(3)_c\times U(1)_{B-L}$, and
the accompanying monopoles are inflated away.

In this brane model, ${\bf 16}_H$, ${\bf \overline{16}}_H$ on B1, and
${\bf 1}$, ${\bf 1'}$ on B2 play the role of $\phi$, $\bar{\phi}$
in Eq.~(\ref{simplepot}).
With the (localized) VEVs of the scalar components of ${\bf 16}_H$,
${\bf \overline{16}}_H$ along the $SU(5)$ singlet direction
(i.e. $\langle\nu^c_H\rangle$, $\langle\overline{\nu}^c_H\rangle$) at B1
after inflation, the $SO(10)$ gauge symmetry breaks to $SU(5)$.
On the other hand, at B2 only the singlets
${\bf 1}$, ${\bf 1'}$ rather than
$H^c$, $\overline{H}^c$ develop VEVs at the minimum of the potential.
Since $\Sigma$ becomes heavy by VEVs of
${\bf 16}_H$, ${\bf \overline{16}}_H$ on B1,
the VEV $\langle\Sigma\rangle$ vanishes after inflation, and so
the symmetry $SU(4)_c\times SU(2)_L\times SU(2)_R$ on B2 is restored.
Consequently, the effective low energy theory after inflation
is the desired MSSM
($=\{SU(5)\}\cap \{SU(4)_c\times SU(2)_L\times SU(2)_R\}$).
We note that the symmetry breaking process
$SU(3)_c\times U(1)_{B-L}\times SU(2)_L\times SU(2)_R\rightarrow
SU(3)_c\times SU(2)_L\times U(1)_{Y}$ does not create any unwanted
topological defects such as monopoles, and so
we have formulated a realistic 5D model in which the monopole problem
is solved without introducing non-renormalizable terms.

While the Goldsotne fields ${\bf (\overline{3},1)}_{-2/3}^{++}$,
${\bf (1,1)}_{-1}^{++}$
(also ${\bf (3,1)}_{2/3}^{++}$, ${\bf (1,1)}_{1}^{++}$)
of $H^c$ ($\overline{H}^c$) are absorbed by the appropriate gauge bosons,
the superhiggs mechanism leaves intact the massless supermultiplets
${\bf (\overline{3},1)}_{1/3}$, ${\bf (3,1)}_{-1/3}$,
which can acquire masses of order $m_{3/2}$
from their couplings to $\langle S\rangle$ after SUSY breaking.
To eliminate this pair from the low energy theory,
we can introduce on B1 a 10-plet with couplings
${\bf 16}_H{\bf 16}_H{\bf 10}$ and
${\bf \overline{16}}_H{\bf \overline{16}}_H{\bf 10}$ (thus,
${\bf 10}$ has an $R$-charge of unity),
and/or a ${\bf (6,1,1)}$ field ($\equiv D$) on B2 with couplings
$H^cH^cD$ and $\overline{H}^c\overline{H}^cD$.
Then, the pair acquires superheavy masses proportional to
$\langle\nu^c_H\rangle$ or $\langle\overline{\nu}^c_H\rangle$, and
the low energy spectrum is precisely the MSSM one.

Note that we introduced the Higgs 16-plets in the bulk rather than
on the $SO(10)$ brane B1 in order to avoid unwanted states
associated with the pseudo-Goldstone symmetry of the superpotential.
Recall that the orbifold compactification breaks $SO(10)$ down
to $SU(4)_c\times SU(2)_L\times SU(2)_R$.

To resolve the DT splitting problem, the Higgs 10-plet ( or ${\bf (1,2,2)}$)
should be introduced
in the bulk (on B2).
By suitable $Z_2\times Z_2'$ parity assignments, the MSSM
Higgs doublets are kept light, while the color triplets become superheavy.

\section{${\bf SU(3)_c\times SU(2)_L\times U(1)_Y\times U(1)_X}$
Model}

In this secstion, we construct the other inflationary model based on
$SU(3)_c\times SU(2)_L\times U(1)_Y\times U(1)_X$.  
We introduce a $U(1)_{PQ}$ axion symmetry and
`matter' parity $Z_2^m$~\cite{khalil}.
For simplicity, let us assume that the MSSM matter superfields as well as
the right-handed neutrinos are brane fields residing at B1.\footnote{If
the first two quark and lepton families reside on B2 where
$SU(5)'\times U(1)_X'$ is preserved, undesirable mass relations between
the down-type quarks and the charged leptons do not arise.
Mixings between the first two and the third families
can be generated by introducing bulk superheavy hypermultiplets
in the spinor representations of $SO(10)$~\cite{5dso10}.  }
They belong in ${\bf 10}_i$, ${\bf \overline{5}}_i$,
and ${\bf 1}_i$ of $SU(5)$, where $i$ is the family index.
Their assigned $U(1)_X$, $U(1)_R$ and $U(1)_{PQ}$ charges and matter parities
appear in Table IX.
\vskip 0.6cm
\begin{center}
\begin{tabular}{|c||c|c|c|c|c|c|c|c|} \hline
Fields & $S$& $~N_H~$& $~\overline{N}_H~$ &$~{\bf 10}_B^{(')}~$&
$~{\bf \overline{10}}_B^{(')}~$ & ${\bf 1}_i$  &
$~~{\bf \overline{5}}_i~~$ & $~{\bf 10}_i~$
\\ \hline \hline
$X^{(')}$ & $0$& $5$& $-5$& $-4$& $4$& $5$& $-3$& $1$
\\
$R$ & $1$ & $0$& $0$& $0$ & $0$ & $1/2$ & $1/2$ & $1/2$
\\
$PQ$ & $0$& $0$& $0$& $0$ & $0$ & $0$ & $-1$ & $-1/2$
\\
$Z_2^m$ & $+$& $+$& $+$& $-$& $-$& $-$& $-$& $-$
\\
\hline \hline
Fields & ~$h_1^{(')}$~ & $h_2^{(')}$ & $~\overline{h}_1^{(')}~$ &
$~\overline{h}_2^{(')}~$ & $\Sigma_1$ & $\Sigma_2$ & $\overline{\Sigma}_1$ &
$\overline{\Sigma}_2$
\\
\hline \hline
$X^{(')}$& $-2$& $-2$& $2$ & $2$ & $0$ & $0$& $0$ & $0$
\\
$R$ & $0$& $1$& $0$ & $1$& $1/2$ & $1/2$ & $0$ & $0$
\\
$PQ$  & $1$& $1$& $3/2$ & $3/2$& $-1$ & $-3/2$ & $1$ & $3/2$
\\
$Z_2^m$& $+$& $-$& $+$& $-$& $+$& $+$& $+$& $+$
\\
\hline
\end{tabular}
\vskip 0.4cm
{\bf Table IX.~}$U(1)_X^{(')}$, $U(1)_R$, $U(1)_{PQ}$ charges
and matter parities of the superfields.
\end{center}

We introduce two pairs of hypermultiplets
$(H_{\bf 10},H^c_{\bf 10^c})$ and
$(H_{\bf \overline{10}},H^c_{\bf \overline{10}^c})$
($=(H_{\bf 10},H^c_{\bf 10^c})$) in the bulk.
The two $SU(5)$ Higgs multiplets $h_1$ and $\overline{h}_1$ (${\bf 5}$ and
${\bf \overline{5}}$) arise from $H_{\bf 10}$ and $H_{\bf \overline{10}}$,
and their $U(1)_R$ charges are chosen to be zero.
As discussed in section 2.2,  
the $N=2$ superpartners $H_{\bf 10}^c$ and $H_{\bf \overline{10}}^c$
also provide superfields $h_2$ and $\overline{h}_2$
with ${\bf 5}^{(')}$ and ${\bf \overline{5}}^{(')}$ representations at B1 (B2).
However, their $U(1)_R$ charges are unity unlike $h_1$ and $\overline{h}_1$.
To make them superheavy
we can introduce another pair of ${\bf 5}$ and ${\bf \overline{5}}$
with zero $U(1)_R$ charges and `$-$' matter parities on the brane.

The superpotential at B1, neglecting the superheavy particles' contributions
except for the inflatons, is given by
\begin{eqnarray} \label{3211}
W&=&\kappa S\bigg(N_H\overline{N}_H
%
-M^2\bigg)
+\frac{\sigma_1}{M_P}\Sigma_1\Sigma_2h_1\overline{h}_1
+\frac{\sigma_2}{M_P}\Sigma_1\Sigma_2\overline{\Sigma}_1\overline{\Sigma}_2
 \\
&&+y^{(d)}_{ij}{\bf 10}_i{\bf 10}_jh_1+
y^{(ul)}_{ij}{\bf 10}_i{\bf \overline{5}}_j\overline{h}_1
+y^{(n)}_{ij}{\bf 1}_i{\bf \overline{5}}_jh_1
+\frac{y^{(m)}_{ij}}{M_P}{\bf 1}_i{\bf 1}_j\overline{N}_H\overline{N}_H ~,
 \nonumber
\end{eqnarray}
where $S$, $N_H$, $\overline{N}_H$, $\Sigma_{1,2}$ and
$\overline{\Sigma}_{1,2}$ are singlet fields.
Their assigned quantum numbers appear in Table IX.
While $S$, $N_H$, $\overline{N}_H$, and $h_1$, $\overline{h}_1$ are
bulk fields, the rest are brane fields residing on B1.
$N_H$, $\overline{N}_H$ should be embedded
in ${\bf 16}_H$, ${\bf \overline{16}}_H$, and
the other components in ${\bf 16}_H$, ${\bf \overline{16}}_H$ could
be made heavy by pairing them with proper brane fields.
From Eq.~(\ref{3211}),
it is straightforward to show that the
SUSY vacuum corresponds to $\langle S\rangle=0$, while
$N_H$ and $\overline{N}_H$ develop VEVs of order $M$.
They break $SU(3)_c\times SU(2)_L\times U(1)_Y\times U(1)_X$
to the MSSM gauge group, and make the massless modes
in $\Sigma_{{\bf 10}_{-4}}$ ($\equiv {\bf 10}_{B}$) and
$\Sigma_{{\bf \overline{10}}_{4}}$ ($\equiv {\bf \overline{10}}_B$)
superheavy~\cite{5dso10}.
From the last term in Eq.~(\ref{3211}), the VEV of $\overline{N}_H$
also provides masses to the right-handed Majorana neutrinos.
Then the `$\tau$' neutrino mass becomes in eV range via the seesaw mechanism. 

Because of the presence of the soft terms,
$\Sigma_{1,2}$ and $\overline{\Sigma}_{1,2}$, which carry $U(1)_{PQ}$ charges,
can obtain intermediate scale VEVs of order $\sqrt{m_{3/2}M_P}$.
They lead to a $\mu$ term of order $m_{3/2}$ in MSSM
as desired \cite{mu}.
Of course, the presence of $U(1)_{PQ}$ also resolves
the strong CP problem~\cite{axion}.
As a result of the $U(1)_{PQ}$ symmetry breaking at the intermediate scale,
there exists a very light axion solving the strong CP problem
in the model \cite{axion}. 

The Higgs fields $h_1$ and $\overline{h}_1$ contain color triplets
as well as weak doublets.
Since the triplets in $h_1$ and $\overline{h}_1$ are just superheavy KK modes,
a small coefficient ($\mu\sim$ TeV) accompanying $h_1\overline{h}_1$
more than adequately suppresses dimension 5 operators that induce proton decay.
Proton decay can still proceed via superheavy gauge bosons with masses
$\approx \pi/y_c$ and are adequately suppressed ($\tau_p\sim 10^{34-36}$ yrs).

Soft SUSY breaking effect and instanton effect
break $U(1)_R$  explicitly to $Z_2$.
Then odd parities under the subgroup $Z_2$ are assigned to the quark and
lepton sector fields and $\Sigma_{1,2}$, $\overline{\Sigma}_{1,2}$.
Since $\Sigma$ fields get VEV, this discrete
symmetry is spontaneously broken and domain walls are created.
Therefore, in this model, we must assume that
the $U(1)_{PQ}$ breaking takes place before or during inflation
so that the induced domain walls are washed out.
At low energy, remaining symmetry is SM gauge group$\times Z_2^{mp}$.

\section{Conclusion}

We have taken the approach that a satisfactory inflationary
scenario should:

\noindent (i) resolve the flatness and horizon problem;

\noindent (ii) resolve cosmological problems associated with topological
defects;

\noindent (iii) give rise to the observed $\delta T/T$ fluctuations;

\noindent (iv) provide a satisfactory explanation of the origin
of the observed baryon asymmetry;

\noindent (v) be well grounded in particle physics.  

\noindent While four dimensional $SO(10)$ models of inflation are hard
to construct, especially if a resolution of DT splitting problem is
also desired, things are much easier if we consider five dimensional $SO(10)$.
The DT splitting problem in $SO(10)$ can be simply resolved
through the 5D orbifold symmetry breaking process.
We find a class of models
in which $\delta T/T\propto (M/M_{\rm Planck})^2$,
where $M$ denotes the symmetry breaking scale associated with inflation,
the scalar spectral index $n_s =0.98\pm 0.01$, $dn_s/d{\rm ln}k\sim 10^{-3}$,
the tensor-to-scalar ratio $r\sim 10^{-8}$, and baryogenesis occurs
via leptogenesis.
We have also shown how the 5D model avoids the monopole problem.  

\vskip 0.3cm
\noindent {\bf Acknowledgments}

\noindent
We were supported for traveling expenses 
to participate in the workshop  
by NSF under contract number INT--9907570.

\end{document}